\documentclass[11pt, a4paper]{article}
\usepackage{jheppub}
\usepackage{latexsym}
\usepackage{pdflscape}
\usepackage{bm}
\numberwithin{equation}{section}
\allowdisplaybreaks

\allowdisplaybreaks[2]

\def\cO{\mathcal{O}}

\def\mint{\int_{-\infty}^\infty\!\cdots\!\int_{-\infty}^\infty}

\newcommand{\be}{\begin{equation}}
\newcommand{\ee}{\end{equation}}
\newcommand{\ba}{\begin{aligned}}
\newcommand{\ea}{\end{aligned}}

\DeclareMathOperator{\Li}{Li}
\DeclareMathOperator{\Ai}{Ai}

\def\({\left(}
\def\){\right)}

\newcommand{\pd}{\partial}

\DeclareMathOperator{\Tr}{Tr}

\newcommand{\nn}{\nonumber \\}

\newcommand{\qu}{\frac{1}{4}}

\def\h#1{\widehat{#1}}

\def\rt#1{\sqrt{#1}}
\def\sitarel#1#2{\mathrel{\mathop{\kern0pt #1}\limits_{#2}}}

\preprint{DESY 14-126}

\title{Probing non-perturbative effects in M-theory}

\author[a]{Yasuyuki Hatsuda}
\author[b]{and Kazumi Okuyama}

\affiliation[a]{DESY Theory Group, DESY Hamburg, \\
Notkestrasse 85, D-22603 Hamburg, Germany}
\affiliation[b]{Department of Physics, \\
Shinshu University, Matsumoto 390-8621, Japan}

\emailAdd{yasuyuki.hatsuda@desy.de} 
\emailAdd{kazumi@azusa.shinshu-u.ac.jp}

\abstract{
The AdS/CFT correspondence enables us to probe M-theory on various backgrounds
from the corresponding dual gauge theories.
Here we investigate in detail a three-dimensional $U(N)$ $\mathcal{N}=4$ super Yang-Mills theory
coupled to one adjoint hypermultiplet and $N_f$ fundamental hypermultiplets,
which is large $N$ dual to M-theory on $AdS_4 \times S^7/\mathbb{Z}_{N_f}$.
Using the localization and the Fermi-gas formulation, we explore non-perturbative corrections
to the partition function.
As in the ABJM theory, we find that there exists a non-trivial pole cancellation mechanism, 
which 
guarantees the theory to be well-defined,
between worldsheet instantons and membrane instantons for all rational (in particular, physical or integral) values of $N_f$.  
}

\begin{document}

\maketitle

\renewcommand{\thefootnote}{\arabic{footnote}}
\setcounter{footnote}{0}
\setcounter{section}{0}

\section{Introduction}\label{sec:intro}
In this paper, we study non-perturbative aspects of M-theory via the AdS/CFT correspondence \cite{Maldacena:1997re}.
Our analysis here is based on the belief that the AdS/CFT correspondence 
(or more generally, the gauge/gravity duality) is exactly true even at quantum level.
This means that gauge theories, if they have gravity duals, provide us a ``non-perturbative definition''
of their dual string theories/M-theory on the corresponding backgrounds.
Recent remarkable developments on exact understandings of gauge theories enable us to
probe the non-perturbative effects in the dual string theories/M-theory, quantitatively.

We are interested in the low energy effective theories on multiple M2-branes, which have
a dual M-theory description on some $AdS_4$ background.
The most well-known example is a $U(N)_k \times U(N)_{-k}$ $\mathcal{N}=6$ supersymmetric 
Chern-Simons-matter
theory, known as the ABJM theory \cite{ABJM}.
The ABJM theory describes the low energy effective theory on the $N$ M2-branes probing a $\mathbb{C}^4/\mathbb{Z}_k$ singularity,
and it is dual to M-theory on $AdS_4 \times S^7/\mathbb{Z}_k$ in the large $N$ limit.
In this paper, we pick up another example: a three-dimensional $U(N)$ $\mathcal{N}=4$ super Yang-Mills theory
coupled to one adjoint hypermultiplet and $N_f$ fundamental hypermultiplets.
This theory also describes the theory on $N$ M2-branes probing a $\mathbb{C}^2 \times \mathbb{C}^2/\mathbb{Z}_{N_f}$ singularity \cite{Benini:2009qs}.
From the Type IIA viewpoint, this is the worldvolume theory
on $N$ D2-branes in the presence of $N_f$ D6-branes.
This theory is dual to M-theory on $AdS_4 \times S^7/\mathbb{Z}_{N_f}$ in the limit
$N \to \infty$ with $N_f$ fixed. Note that the $\mathbb{Z}_{N_f}$ quotient on
$S^7$ acts
differently from the ABJM case. 
We can probe M-theory from these theories via the AdS/CFT correspondence.
We would like to find universal (background independent) properties in M-theory
through various examples.

As shown in \cite{KWY1, Jafferis:2010un, Hama:2010av}, in many supersymmetric gauge theories on $S^3$, infinite dimensional path integrals
for the partition function and vacuum expectation values (VEVs) of BPS Wilson loops reduce to finite dimensional matrix integrals
by using the localization technique.
Due to this drastic simplification, one can, in principle, evaluate the partition function and the Wilson loop VEVs
beyond the perturbation theory.
However, it is still non-trivial to extract their large $N$ behaviors from the matrix integrals.
The traditional matrix model technique is very helpful in the analysis in the 't Hooft limit.
In the ABJM matrix model, a systematic analysis in the 't Hooft limit was done in \cite{MP1, DMP1}.
However it is not easy to access the M-theory regime in this way (see \cite{Herzog:2010hf}).
It is desirable to find more efficient ways to understand the M-theory regime systematically.

Recently, Mari\~no and Putrov proposed a very interesting formulation, known as the 
Fermi-gas approach, to analyze matrix models
for a wide class of 3d Chern-Simons-matter theories \cite{MP2}.
This Fermi-gas approach was successfully applied to the ABJM theory
and revealed a very detailed structure of the 
non-perturbative effects in M-theory.
It turned out that the existence of two types of instantons, 
i.e. worldsheet instantons and membrane instantons, 
is crucial for the non-perturbatively complete definition of the theory.
In particular, the worldsheet instanton correction diverges at every physical value of
the coupling, and that divergence is precisely canceled by the
similar,  but opposite sign, divergence of the membrane instanton correction \cite{HMO2}.
This pole cancellation mechanism is conceptually very important, since this mechanism guarantees 
that we can go smoothly from the weak coupling (Type IIA) regime to the strong coupling
(M-theory) regime.
More practically, this mechanism gives strong constraint on the possible form
of membrane instantons. For the ABJM case, we can actually find 
the analytic form of a first few membrane instanton coefficients using 
this pole cancellation condition, 
together with some other input from the semi-classical expansion of Fermi-gas \cite{HMO2,CM,HMO3}.
Based on these analytic results, it was finally found in \cite{HMMO, KM} that the membrane instantons in 
the ABJM theory are completely determined by the refined topological string on local 
$\mathbb{P}^1\times\mathbb{P}^1$, 
in the Nekrasov-Shatashvili limit \cite{Nekrasov:2009rc}, while
the worldsheet instantons are given by the standard topological string on the same
manifold \cite{DMP1, HMO2}.
In addition, there are bound states of membrane instantons and worldsheet instantons,
whose contributions are finally absorbed into the worldsheet instanton corrections by
the effective shift of chemical potential of the Fermi-gas system \cite{HMO3}.
In the ABJ theory \cite{ABJ}, the similar structure was also found \cite{MM,HO,Kallen:2014lsa}
based on the results \cite{Awata:2012jb, Honda:2013pea}.

In the present paper, we will study the $S^3$
partition function $Z(N_f,N)$ of the $U(N)$ $\mathcal{N}=4$ super Yang-Mills theory
coupled to one adjoint hypermultiplet and $N_f$ fundamental hypermultiplets.
Using the localization technique, the computation of
$Z(N_f,N)$ boils down to a matrix integral, which was named as the 
$N_f$ matrix model in \cite{GM}.
It is known that the grand partition function of
the $N_f$ matrix model can be recast as a Fermi-gas system,
and some of its properties were studied in \cite{Mezei:2013gqa,GM}.
This $N_f$ matrix model is an interesting first step beyond ABJ(M) theory to 
study the non-perturbative effects in M-theory.
However, it turned out that it is not straightforward to 
apply the strategy in the previous paragraph, which was successful in the ABJM case \cite{HMO2},
to the $N_f$ matrix model:
\begin{enumerate}
 \item Find the worldsheet instanton coefficients and their pole structure. 
 \item Determine the analytic form of the membrane instanton correction by combining the small $N_f$ expansion and the pole cancellation condition.
\end{enumerate}
At the step 1, in the case of  ABJM theory,
the analytic form of the worldsheet instanton correction is available  thanks
to the relation to the topological string on local $\mathbb{P}^1\times\mathbb{P}^1$.
On the other hand, the $N_f$ matrix model does not seem to have a direct connection 
to the topological string
theory, and hence the analytic form of the worldsheet instanton correction is not known,
except for the genus zero part \cite{GM}.
Currently, there is no systematic way to compute the worldsheet instanton corrections as analytic functions of $N_f$.
To overcome this problem, we first compute the exact values of the partition function $Z(N_f,N)$
for various integral values of $N_f$ up to some high $N$, and then guess the worldsheet instanton 
coefficients as functions of $N_f$ using the exact data of $Z(N_f,N)$.
In this way, we indeed 
find the analytic forms of worldsheet instanton coefficients up to three-instanton \eqref{eq:WSinst}.
We note that our conjecture gives an all-genus prediction in each instanton sector 
when taking the 't Hooft limit.
Our conjecture passes many non-trivial checks.

The small $N_f$ expansion 
at step 2 is also difficult to be carried out, 
since the density matrix (or Hamiltonian) of the Fermi-gas explicitly depends on $N_f$ 
\cite{Mezei:2013gqa,GM}.
Nevertheless, 
we successfully find the first few terms of the small $N_f$ expansion of the grand potential
by analyzing the so-called thermodynamic Bethe ansatz (TBA) equations. 
Combining the small $N_f$ expansion and the pole cancellation condition, we determine
the membrane one-instanton coefficient completely \eqref{eq:b1}
and find a part of the
membrane two-instanton coefficients \eqref{eq:a2}, as analytic functions of $N_f$.
As a non-trivial check, 
we show that our conjecture of membrane instantons
is consistent with the numerical solution of the TBA equations for $0<N_f<1$.
These results clearly show that the pole cancellation mechanism is a general phenomenon
in M-theory, not the special property of the ABJ(M) theory.

This paper is organized as follows: In section 2, we review the known properties of the $N_f$
matrix model, including the Fermi-gas approach, the TBA equations, 
and the 't Hooft and the M-theory limits of this model.
In section 3, first we explain our algorithm to compute the exact values of the partition functions
$Z(N_f,N)$ for various integral values of $N_f$. Then, using these exact values,
we analyze the structure of the grand potential. We find that the constant $A(N_f)$ in
the grand potential is related to the constant map contribution of the topological string.
We also determine the analytic forms of the worldsheet instanton coefficients up to three-instanton.
In section 4, we consider the membrane instanton corrections. First we study the small
$N_f$ expansion of the TBA equations, then we consider the 
pole cancellation condition between the worldsheet instantons and the membrane instantons.
Finally, combining these two inputs, we determine the analytic forms of the membrane one-instanton and 
of a part of the membrane two-instanton. Section 5 is the conclusion.
We also have three appendices A, B, and C, summarizing some results used in the main text.

\section{Review of  Fermi-gas approach}\label{sec:review}
Let us start by reviewing the exact computation of the partition function on $S^3$ by using the localization \cite{KWY1}
and the Fermi-gas approach \cite{MP2}.
The localization reduces the partition function to a finite dimensional matrix integral. 
We rewrite this matrix integral as a partition function of certain one-dimensional ideal Fermi-gas system as in \cite{MP2}.
This approach is quite powerful, and allows us to analyze the non-perturbative corrections to the partition function.

\subsection{From matrix model to Fermi-gas}
\begin{figure}[tb]
\begin{center}
\begin{tabular}{cc}
\resizebox{55mm}{!}{\includegraphics{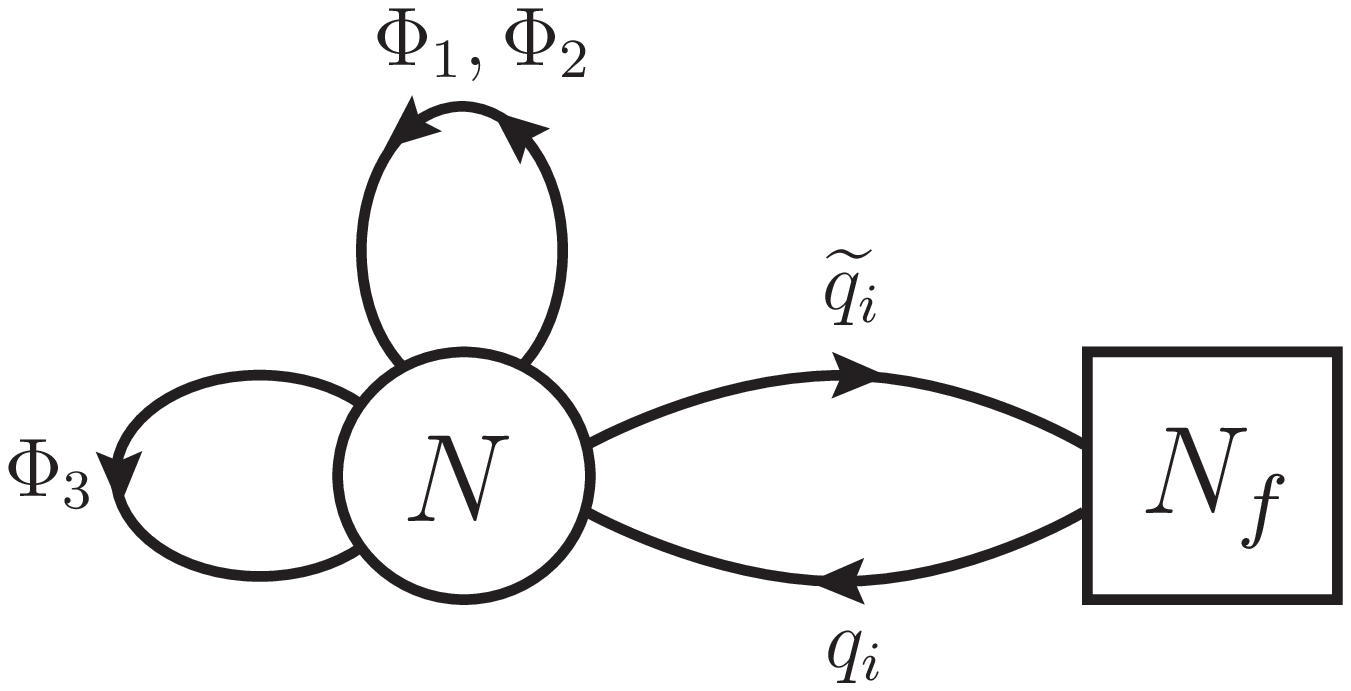}}
\hspace{15mm}
&
\resizebox{40mm}{!}{\includegraphics{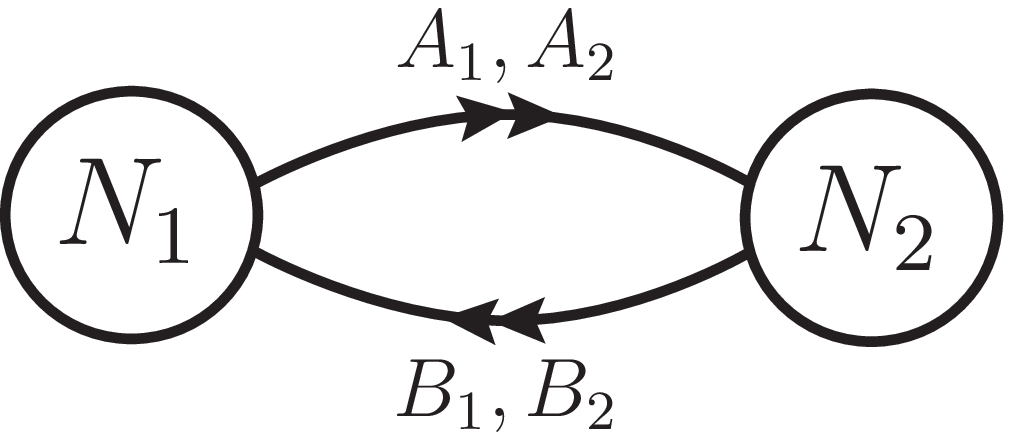}}
\vspace{5mm}
\\ (a)\hspace{5mm}  & (b)
\end{tabular}
\end{center}
  \caption{(a) The quiver diagram for the $U(N)$ $\mathcal{N}=4$ SYM
with an adjoint and $N_f$ fundamental hypermultiplets.
We use $\mathcal{N}=2$ notation.
The adjoint hypermultiplet consists of two adjoint chiral multiplets $(\Phi_1,\Phi_2)$, and $\Phi_3$ is the chiral multiplet
that forms the $\mathcal{N}=4$ vector multiplet together with the $\mathcal{N}=2$ vector multiplet.
(b) The quiver for the $U(N_1)_k \times U(N_2)_{-k}$ ABJ(M) theory. 
The partition function of the former with $N_f=1,2$ is related to that of the latter at $k=1,2$, respectively.}
  \label{fig:quiver}
\end{figure}

In this paper we consider the $U(N)$ $\mathcal{N}=4$ super Yang-Mills theory on $S^3$
with one adjoint hypermultiplet and $N_f$ fundamental hypermultiplets.
The quiver diagram for this theory is shown in figure~\ref{fig:quiver}(a).
Following the general localization procedure in \cite{KWY1}, one can immediately write down the partition function
of this theory, and the result is given by the matrix integral
\be
Z(N_f,N)=\frac{1}{N!} \int \prod_{j=1}^N \frac{dx_j}{4\pi} \frac{1}{(2\cosh \frac{x_j}{2})^{N_f}}
\prod_{j<k} \tanh^2 \( \frac{x_j-x_k}{2} \).
\label{eq:Z}
\ee
Our goal in this paper is to understand the large $N$ behavior of this model, 
the so-called $N_f$ matrix model, including non-perturbative corrections.
We note that the matrix integral \eqref{eq:Z} can be evaluated for arbitrary value of $N_f$ while
the physical theory is defined only for integral $N_f$.
It is well-known that the theory with $N_f=1$ is equivalent to the ABJM theory at Chern-Simons level $k=1$ 
via the mirror symmetry \cite{ABJM}.
The equality of the partition functions of both theories was directly shown in \cite{KWY2}. 
Interestingly, the partition function for $N_f=2$ is also related to the partition function of the $U(N)_k \times U(N+1)_{-k}$ 
ABJ theory at $k=2$
\begin{align}
 |Z_{\rm ABJ}(k=2,N,N+1)|=\frac{1}{\sqrt{2}}Z(N_f=2,N),
\end{align}
where the factor $1/\sqrt{2}$ comes from the partition function of $U(1)_{k=2}$
pure Chern-Simons theory.
This relation can be checked by rewriting the ABJ partition function as in \cite{HO}.
The quiver diagram of the ABJ(M) theory is shown in figure \ref{fig:quiver}(b).

It is not easy to perform the matrix integral \eqref{eq:Z} directly.%
\footnote{One interesting approach is to evaluate the multi-integral numerically by using the Monte Carlo method
as in the (mirror) ABJM matrix model \cite{KEK}.} 
However, there is a very efficient way to compute the exact values of the partition function.
This method was first proposed in \cite{MP2} (see also \cite{Okuyama:2011su}), and is now called the Fermi-gas approach.
The key idea is to rewrite the partition function \eqref{eq:Z} as
\be
Z(N_f,N)=\frac{1}{N!}\sum_{\sigma \in S_N}(-1)^\sigma \int \! d^N x \prod_{j=1}^N \rho(x_j,x_{\sigma(j)}),
\label{eq:Z-FG}
\ee
where
\be
\rho(x_1,x_2)=\frac{1}{2\pi} \frac{1}{(2\cosh \frac{x_1}{2})^{N_f/2}} \frac{1}{(2\cosh \frac{x_2}{2})^{N_f/2}}
 \frac{1}{2\cosh \( \frac{x_1-x_2}{2}\)}.
\ee
To derive \eqref{eq:Z-FG}, we used the Cauchy determinant formula.
The partition function \eqref{eq:Z-FG} can be interpreted as the partition function of an ideal
Fermi-gas system described by the density matrix $\rho$ \cite{MP2}.
In the following analysis, it is very convenient to introduce the grand canonical partition function and 
the grand potential
\be
\Xi(N_f,z)=1+\sum_{N=1}^\infty z^N Z(N_f,N), \qquad J(N_f,\mu)=\log \Xi (N_f,z),
\ee
where $z=e^{\mu}$ is a fugacity with a chemical potential $\mu$ in the grand canonical ensemble.
As was discussed in \cite{MP2}, for the partition function with the form \eqref{eq:Z-FG}, the grand partition function is written
as a Fredholm determinant%
\footnote{We have used the well-known identity: $\log \det X = \Tr \log X$ for a matrix $X$.}
\be
\Xi(N_f,z)=\det (1+ z \rho)=\exp \biggl[ -\sum_{n=1}^\infty \frac{(-z)^n}{n} \Tr \rho^n \biggr],
\ee
where the trace of $\rho$ is defined by
\be
\Tr \rho^n =\int_{-\infty}^\infty dx_1\cdots dx_n \, \rho(x_1,x_2)\rho(x_2,x_3) \cdots \rho(x_n,x_1).
\ee
Therefore the basic problem is how to compute $\Tr \rho^n$.
This is still not easy, but as will be seen in the next section, we can 
compute it recursively.
Once we know the grand potential $J(N_f,\mu)$, it is easy to reconstruct the canonical partition function by
\be
Z(N_f,N)=\int_{-\pi i}^{\pi i} \frac{d\mu}{2\pi i} e^{J(N_f,\mu)-N \mu}.
\label{eq:Z-J}
\ee

\subsection{TBA equations}
The density matrix $\rho$ takes the form
\be
\rho(x_1,x_2)=\frac{1}{2\pi} \frac{e^{-\frac{1}{2}U(x_1)-\frac{1}{2}U(x_2)}}{2\cosh \( \frac{x_1-x_2}{2} \)},\qquad
U(x)=N_f \log \left[ 2\cosh \frac{x}{2} \right].
\ee
Interestingly, for the kernel with this form, we can compute the grand potential from the TBA equations \cite{Zamolodchikov:1994uw}.
The TBA integral equations are given by
\be
\ba
\log R_+(x)&=-U(x)+\int_{-\infty}^\infty \frac{dx'}{2\pi} \frac{\log(1+\eta^2(x'))}{\cosh (x-x')},\\
\eta(x)&=-z \int_{-\infty}^\infty \frac{dx'}{2\pi} \frac{R_+(x')}{\cosh(x-x')}.
\ea
\label{eq:TBA1}
\ee
For the solutions of these equations, we also define
\be
R_-(x)=R_+(x) \int_{-\infty}^\infty \frac{dx'}{\pi} \frac{\arctan \eta(x')}{\cosh^2(x-x')}.
\label{eq:TBA2}
\ee
Once these functions are determined, the grand potential is computed by
\be
\frac{\pd J_\pm}{\pd z}=\frac{1}{4\pi} \int_{-\infty}^\infty dx \, R_\pm (x),\qquad
J_\pm(z)=\frac{J(z) \pm J(-z)}{2}.
\label{eq:Jpm-TBA}
\ee
We note that the integral equations \eqref{eq:TBA1} and \eqref{eq:TBA2} can be recast as the following functional relations \cite{CM}, 
called the Y-system in the literature,
\be
\ba
R_+\( x+\frac{\pi i}{2} \) R_+\(x- \frac{\pi i}{2} \) \exp \left[ U\( x+\frac{\pi i}{2} \)+U\( x-\frac{\pi i}{2} \) \right]&=1+\eta^2(x),\\
\eta\( x+\frac{\pi i}{2} \)+\eta\( x-\frac{\pi i}{2} \)&=-z R_+(x),
\ea
\label{eq:FR1}
\ee
and
\be
\frac{R_-(x+\frac{\pi i}{2} )}{R_+ ( x+\frac{\pi i}{2} )}
-\frac{R_-(x-\frac{\pi i}{2} )}{R_+ ( x-\frac{\pi i}{2} )}
=2i \frac{\eta'(x)}{1+\eta^2 (x)}.
\label{eq:FR2}
\ee
The TBA equations \eqref{eq:TBA1} and \eqref{eq:TBA2} are powerful in the numerical computation for various
values of $N_f$ as in \cite{HMO1}.
The functional equations \eqref{eq:FR1} and \eqref{eq:FR2}, on the other hand, are useful 
in the semi-classical analysis \cite{CM}.
The TBA equations for the ABJM case were studied in \cite{HMO1, PY, CM}.

\subsection{'t Hooft limit and M-theory limit}
We want to understand the large $N$ behavior of the partition function \eqref{eq:Z} (or equivalently \eqref{eq:Z-FG}).
There are two interesting limits.
One is the standard 't Hooft limit, in which the parameters are taken as follows:
\be
N \to \infty ,\qquad N_f \to \infty ,\qquad \lambda = \frac{N}{N_f} \text{ : fixed},
\ee
where $\lambda$ is the 't Hooft coupling.
In this limit, the free energy admits the perturbative genus expansion (plus non-perturbative contribution)
\be
F(g_s, \lambda)=-\log Z(N_f, N)=\sum_{g=0}^\infty g_s^{2g-2} F_g(\lambda)+F_\text{np}(g_s, \lambda), \qquad g_s=\frac{1}{N_f},
\ee
where $F_g(\lambda)$ is the genus $g$ contribution, and $F_\text{np}(g_s, \lambda)$ is the non-perturbative
correction in $g_s$.
In the ABJM matrix model, the genus zero contribution was 
computed by the standard matrix model technique,
and the 
higher 
genus corrections were determined from  
the holomorphic anomaly equations
\cite{MP1, DMP1}.
In the $N_f$ matrix model, the genus zero contribution was computed in \cite{GM}.
However, it is difficult to compute the higher genus correction.%
\footnote{We thank M. Mari\~no for pointing out the difficulty of the higher genus computation in the $N_f$ matrix model.}
The non-perturbative correction $F_\text{np}(g_s, \lambda)$ is also very difficult to be computed 
from the usual matrix model approach.
Some interesting results on the non-perturbative corrections in the ABJM matrix model are found in \cite{DMP2, Grassi:2014cla}.%
\footnote{In the ABJM matrix model, the perturbative genus expansion is very likely Borel summable.
One of the conclusions in \cite{Grassi:2014cla} is that the Borel resummation of the genus expansion 
does not present the exact result, 
and one needs to consider the non-perturbative contribution $F_\text{np}(g_s,\lambda)$. 
This non-perturbative contribution is caused by so-called complex instantons \cite{DMP2, Grassi:2014cla},
and interpreted as D2-brane instanton effects in Type IIA string theory \cite{DMP2}.}
We emphasize that the Fermi-gas approach overcomes this difficulty, and we can predict analytic results for
$F_\text{np}(g_s, \lambda)$.
One important consequence is that the perturbative genus expansion is insufficient in the finite
$g_s$ regime, and the existence of the
non-perturbative contribution $F_\text{np}(g_s, \lambda)$ is essential
for the consistency of the theory.

We note that the genus $g$ contribution $F_g(\lambda)$ at strong coupling contains the non-perturbative corrections in 
$\alpha' (\sim 1/\sqrt{\lambda})$, 
which has the exponentially suppressed contribution (see \cite{GM} for the genus zero contribution)
\be
\cO(e^{-2\pi \sqrt{2\lambda}}).
\label{eq:WS-order}
\ee
From the dual Type IIA string point of view, such corrections are caused by the worldsheet instanton wrapping a two-cycle.
On the other hand, the non-perturbative part $F_\text{np}(g_s, \lambda)$ contains the exponentially suppressed contribution
\be
\cO(e^{-\pi \sqrt{2\lambda}/g_s}).
\label{eq:D2-order}
\ee
As discussed in \cite{DMP2}, such non-perturbative corrections come from the D2-branes wrapping a three-cycle. 
In this paper, we refer these corrections to the membrane instanton corrections because these are purely non-perturbative effects in $g_s$.

The other interesting limit is the following one:
\be
N \to \infty ,\qquad N_f \text{ : fixed},
\ee
corresponding to a direct thermodynamic limit of the Fermi-gas system.
In this limit, the gauge theory is dual to M-theory on $AdS_4 \times S^7/\mathbb{Z}_{N_f}$, and thus
we call this limit as the M-theory limit here.
In the following analysis, we mainly focus on the M-theory limit.
In the M-theory limit, the worldsheet instanton correction \eqref{eq:WS-order} and the membrane instanton correction
\eqref{eq:D2-order} have the same order
\be
\cO(e^{-2 \pi \sqrt{2N/N_f}}),\qquad \cO(e^{-\pi \sqrt{2N_f N}}).
\label{eq:WS-D2-order}
\ee
This is because, in the M-theory regime, both instantons are up-lifted to M2-branes wrapping two different types of three-cycles.
See figure~1 in \cite{HMMO} for more detail.

Before closing this section, let us comment on the large $N$ limit in the grand canonical ensemble.
From the integral transformation \eqref{eq:Z-J}, the partition function can be evaluated by the saddle point approximation 
in the large $N$ limit.
The saddle point equation is given by
\be
J'(\mu_*)-N=0.
\ee
As was discussed in \cite{Mezei:2013gqa, GM}, the grand potential behaves in the large $\mu$ limit as
\be
J(N_f, \mu) \approx \frac{2}{3\pi^2 N_f}\mu^3 \qquad (\mu \to \infty).
\ee
Therefore, the saddle point is given by
\be
\mu_*=\pi \sqrt{\frac{N_f N}{2}}.
\ee
This means that the large $N$ limit in the canonical ensemble corresponds to the large $\mu$
limit in the grand canonical ensemble.
The saddle point analysis presents a simple derivation of the $N^{3/2}$ behavior of the free energy \cite{MP2}
\be
F(N_f,N) \approx -J(N_f, \mu_*)+N \mu_* \approx \frac{\pi \sqrt{2N_f}}{3} N^{3/2} \qquad (N \to \infty). 
\label{eq:F-largeN}
\ee
This reproduces the matrix model result \cite{Mezei:2013gqa}.
As we will see later, the Airy function behavior is also derived easily 
from the grand canonical analysis. 
Finally, the worldsheet instanton correction and the membrane instanton correction
in \eqref{eq:WS-D2-order} correspond to the exponentially suppressed corrections
\be
\cO(e^{-4\mu/N_f}),\qquad \cO(e^{-2\mu}),
\ee
respectively, in the grand canonical ensemble.

\section{Exploring non-perturbative effects}\label{sec:np}
In this section, we investigate the non-perturbative corrections to the partition function \eqref{eq:Z}
in the large $N$ limit.
As explained in the previous section, the large $N$ limit corresponds to the large $\mu$ limit
in the grand canonical ensemble.
We thus concentrate our attention on the large $\mu$ behavior of the grand potential.
We first compute the exact values of the partition function for various integral $N_f$ by using the Fermi-gas approach
developed in \cite{PY,HMO1,HMO2}.
Next, using these exact data, we extract the non-perturbative corrections to the grand potential.
Based on these results, we look for exact forms of the worldsheet instanton corrections for general $N_f$.
The membrane instanton corrections are explored in the next section.

\subsection{Exact computation of the partition function}\label{sec:exact}
In this subsection, we compute the exact values of the partition function for some integral values of $N_f$.
Our strategy is the same as that in the ABJM theory \cite{HMO1, HMO2}.
We first divide the density matrix $\rho$ into parity even/odd part
\be
\rho(x_1,x_2)=\rho_+(x_1,x_2) + \rho_-(x_1,x_2),\quad \rho_\pm(x_1,x_2)=\frac{\rho(x_1,x_2)\pm \rho(x_1,- x_2)}{2}.
\ee
Then, the grand partition function is factorized into two parts \cite{HMO1},
\be
\Xi(z)=\det(1+z \rho_+) \det(1+z \rho_-).
\ee
As in \cite{HMO1, HMO2}, we write $\rho_\pm$ as the forms
\be
\rho_\pm(x, y)= \frac{E_\pm(x)E_\pm(y)}{\cosh x+ \cosh y},
\ee
where
\be
E_+(x)=\frac{\cosh \frac{x}{2}}{(2\cosh \frac{x}{2})^{N_f/2}},\qquad
E_-(x)=\frac{\sinh \frac{x}{2}}{(2\cosh \frac{x}{2})^{N_f/2}}.
\ee
Now we can apply the result in \cite{TW} to the kernels $\rho_\pm$.
The important consequence is that we can compute $\rho_\pm^n$ from
functions with one variable:
\be
\ba
\rho_\pm^{2n+1}(x,y)&=\frac{E_\pm(x)E_\pm(y)}{\cosh x+\cosh y}\sum_{\ell=0}^{2n} (-1)^\ell \phi_\pm^\ell(x) \phi_\pm^{2n-\ell}(y),\\
\rho_\pm^{2n}(x,y)&=\frac{E_\pm(x)E_\pm(y)}{\cosh x-\cosh y}\sum_{\ell=0}^{2n-1} (-1)^\ell \phi_\pm^\ell(x) \phi_\pm^{2n-1-\ell}(y),
\ea
\label{eq:TW}
\ee 
where the functions $\phi_\pm^\ell(x)$ are determined by the following integral equations recursively
\be
\ba
\phi_+^\ell(x)&=\frac{1}{\cosh \frac{x}{2}} \int_{-\infty}^\infty \frac{dx'}{2\pi} \frac{1}{2\cosh\(\frac{x-x'}{2}\)} \frac{\cosh \frac{x'}{2}}{(2\cosh \frac{x'}{2})^{N_f}}
\phi_+^{\ell-1}(x'),\\
\phi_-^\ell(x)&=\frac{1}{\cosh \frac{x}{2}} \int_{-\infty}^\infty \frac{dx'}{2\pi} \frac{1}{2\cosh\(\frac{x-x'}{2}\)} 
\frac{\sinh \frac{x'}{2} \tanh\frac{x'}{2}}{(2\cosh \frac{x'}{2})^{N_f}} \phi_-^{\ell-1}(x'),
\ea
\label{eq:int_eq}
\ee
with the initial conditions $\phi_\pm^0(x)=1$.
For a derivation of these equations, see \cite{HMO1}.
In \cite{HMO1, HMO2}, we further found that the grand partition function is expressed only in terms of the even parity part:
\be
\Xi(z)=\det(1-z^2 \rho_+^2)G(z),\qquad G(z)=\sum_{\ell=0}^\infty \phi_+^\ell(0) z^\ell. 
\label{eq:Xi}
\ee
This result is very useful in the practical computation.
In summary, we first solve the integral equation \eqref{eq:int_eq} for $\phi_+^\ell(x)$ recursively. 
This can be done very efficiently by using the technique in \cite{PY, HMO2}.
We then compute $\Tr \rho_+^{2n}$ to know $\det (1-z^2 \rho_+^2)$ by using \eqref{eq:TW}.
We finally read off the coefficient of $z^N$ in \eqref{eq:Xi} that is nothing but $Z(N_f,N)$.

\begin{figure}[tb]
  \begin{center}
    \includegraphics[width=10cm]{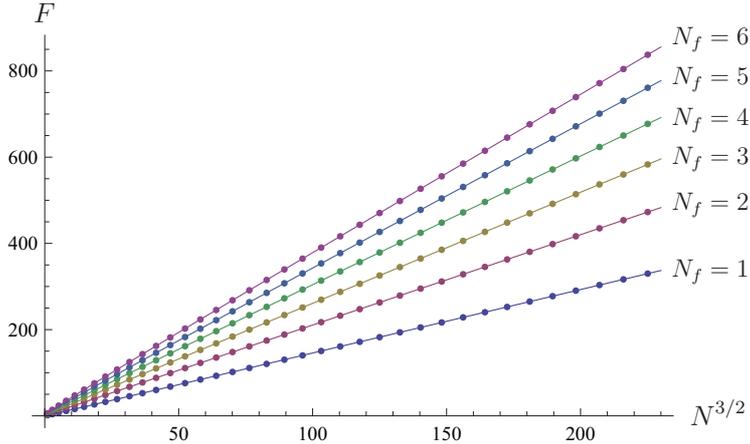}
  \end{center}
  \vspace{-0.5cm}
  \caption{The free energy $F=-\log Z$ for $N_f=1,\dots,6$ as a function of $N^{3/2}$. 
    The dots represent the exact values
    computed by our algorithm while the solid lines represent the perturbative part given by the Airy function \eqref{eq:Zpert}.}
  \label{fig:exactF}
\end{figure}

We have computed the exact values of $Z(N_f,N)$ 
for $N_f=1,2,\dots, 12, 14, 16$,
up to certain values of $N$.
Since the results are very complicated, we cannot write them down here.
Instead, a set of ancillary files for these values readable in {\tt Mathematica}%
\footnote{One can import those files to {\tt Mathematica} by 
{\tt Import["file.dat", "List"]//ToExpression}.} is attached to this paper on {\tt arXiv}.
In figure~\ref{fig:exactF}, we show the free energy $F=-\log Z$ for $N_f=1,2,\dots, 6$ as a function of $N^{3/2}$.
The dots represent the exact values while the solid lines represent the leading Airy function behaviors given by \eqref{eq:Zpert}.%
\footnote{To use this formula, we need the non-trivial function $A(N_f)$. In subsection~\ref{sec:A}, 
we give an exact form of $A(N_f)$. See \eqref{eq:A}.
}
The exact values of the free energy indeed show the $N^{3/2}$ behaviors, 
and also a very good agreement with the Airy function \eqref{eq:Zpert} even for small $N$.

\subsection{General structure of the grand potential}
Next we consider the structure of the grand potential.
In the large $\mu$ limit, the grand potential takes the following form \cite{Mezei:2013gqa, GM}
\be
J(N_f, \mu)=\frac{C(N_f)}{3}\mu^3+B(N_f) \mu+A(N_f)+J_\text{np}(N_f, \mu) \qquad (\mu \to \infty),
\label{eq:J-largemu}
\ee
where
\be
C(N_f)=\frac{2}{\pi^2 N_f},\qquad B(N_f)=\frac{1}{2N_f}-\frac{N_f}{8},
\ee
and $A(N_f)$ is a non-trivial function of $N_f$.
The remaining part $J_\text{np}(N_f,\mu)$ is the exponentially suppressed correction in $\mu \to \infty$.
As we have seen in the previous section, there are two such corrections,
coming from the worldsheet instantons and the membrane instantons.
In addition, there are also ``bound states'' of these two kinds of instantons \cite{CM}.
Taking into account of these bound states, the non-perturbative correction 
has the following expansion%
\footnote{Strictly speaking, the non-perturbative correction $J_\text{np}(N_f,\mu)$ contains an additional contribution
that shows oscillatory behavior \cite{HMO2}. However, this contribution can be removed by deforming the integration contour
in \eqref{eq:Z-J} from $[-\pi i, \pi i]$ to $[-i\infty, i\infty]$. Below, we always take the deformed contour when going back to the canonical
ensemble, thus we can drop this oscillatory contribution.
Note that the oscillatory contribution seems to play an important role in the analysis
of the ``orbifold'' ABJM theory \cite{Honda:2014ica}.} 
\be
J_\text{np}(N_f, \mu)=\sum_{\substack{\ell,m=0\\(\ell,m)\ne (0,0)}}^\infty f_{\ell,m}(N_f,\mu)
\exp \left[ -\(2\ell+\frac{4m}{N_f} \) \mu \right].
\label{eq:Jnp}
\ee
The structure is very similar to the ABJM case \cite{CM},
but the explicit forms of the coefficients look quite different, as
we will see later.
The worldsheet instanton correction corresponds to $\ell=0$,
and the membrane instanton correction to $m=0$.
The others are understood as their bound states.
If we ignore the non-perturbative correction $J_\text{np}(N_f,\mu)$,
the grand potential \eqref{eq:J-largemu} leads to the following canonical partition function \cite{MP2}
\be
Z_\text{pert}(N_f,N)=C(N_f)^{-1/3} e^{A(N_f)} \Ai \left[ C(N_f)^{-1/3}(N-B(N_f)) \right].
\label{eq:Zpert}
\ee
where $\Ai (z)$ is the Airy function.
This result is understood as the all-genus resummation after neglecting all the exponentially suppressed corrections \cite{FHM}.
Using the asymptotic expansion of the Airy function, one can, of course, reproduce the $N^{3/2}$
behavior \eqref{eq:F-largeN} in the large $N$ limit.

Our remaining task is to determine the non-trivial functions $A(N_f)$ and $f_{\ell,m}(N_f,\mu)$.
In the ABJ(M) theory, this program has already been done with the help of an accidental connection to the topological
string on local $\mathbb{P}^1 \times \mathbb{P}^1$ (see \cite{HMO2, CM, HMO3, HMMO, MM, HO}).
However, in our case of the $N_f$ matrix model, 
we do not know a nice connection to the topological string.
Therefore we do not have any guiding principles to determine the non-perturbative corrections systematically, and
it is challenging to understand $f_{\ell,m}(N_f,\mu)$.
To explore the non-perturbative effects, we here take the following strategy:
\begin{itemize}
\item Using the exact data computed in the previous subsection, we extract the non-perturbative corrections for various integral 
values of $N_f$.
\item Based on these data, we conjecture the worldsheet instanton correction $f_{0,m}(N_f,\mu)$
for general $N_f$ order by order.
\item The (conjectured) worldsheet instanton correction diverges for some values of $N_f$.
These singularities must be canceled by the other contributions because the theory is always well-defined.
This pole cancellation mechanism was first found in the ABJM theory \cite{HMO2}.
Using this mechanism, we can determine the pole structure of the membrane instanton correction.
Combining this information with some other inputs, we fix the analytic forms of the membrane instanton corrections.
\item The obtained results can be compared with the numerical results computed from the TBA equations in section~\ref{sec:review} for
various (non-integral) values of $N_f$. This comparison gives a highly non-trivial test of our conjecture. 
\end{itemize}
Of course, the analytic forms of higher instanton corrections become very complicated,
and it gets more and more difficult to determine them in this way.
In this paper, following the above strategy, we indeed determine the worldsheet instanton correction up to $m=3$,
and also the leading membrane instanton correction (and a part of the next-to-leading correction).
So far, we cannot obtain any results on the bound states.
This should be understood in the future work.

\subsection{The constant part}\label{sec:A}
\begin{figure}[tb]
  \begin{center}
    \includegraphics[width=8cm]{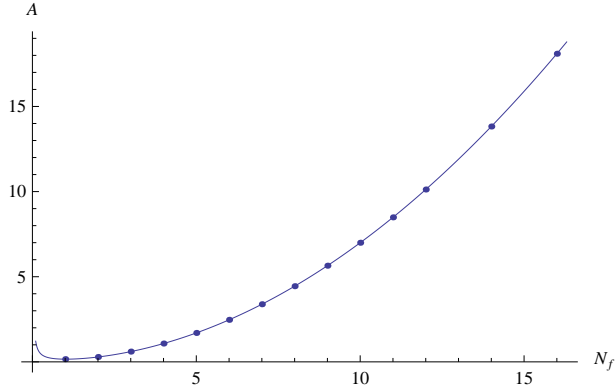}
  \end{center}
  \vspace{-0.5cm}
  \caption{The constant contribution $A(N_f)$. The numerical estimations \eqref{eq:A-est} from exact $Z(N_f,N)$ in the previous section
are shown by the dots, and our conjecture \eqref{eq:A} by the solid line.}
  \label{fig:A}
\end{figure}

Before proceeding to the instanton corrections, we give a conjecture of the exact form of $A(N_f)$.
We find that the constant part $A(N_f)$ is exactly related to that in the ABJM theory as
\be
A(N_f)=\frac{1}{2} \left[ A_\text{const}(N_f)+A_\text{const}(1) N_f^2 \right],
\label{eq:A}
\ee
where $A_\text{const}(k)$ is the constant part appearing in the grand potential in the ABJM Fermi-gas \cite{MP2}.
Although we do not have a proof of this conjecture, it passes many non-trivial tests 
as we will see below.
Note that $A_\text{const}(k)$ corresponds to the constant map contribution in the topological string \cite{KEK}.
The small $k$ expansion of $A_\text{const}(k)$ was first computed in \cite{MP2},
and then the all-loop formula and its integral expression were conjectured in \cite{KEK}.
As derived in appendix~\ref{sec:Aconst}, we find another simpler integral expression of $A_\text{const}(k)$:
\be
A_\text{const}(k)=\frac{2\zeta(3)}{\pi^2 k}\(1-\frac{k^3}{16}\)
+\frac{k^2}{\pi^2} \int_0^\infty dx \frac{x}{e^{k x}-1}\log(1-e^{-2x}).
\label{eq:A-int}
\ee
In particular, we find closed form expressions for any integer $k$,
\be
\ba
A_\text{const}(k)=\begin{cases}
\displaystyle
-\frac{\zeta(3)}{\pi^2 k}-\frac{2}{k} \sum_{m=1}^{\frac{k}{2}-1} m\(\frac{k}{2}-m\) \log \( 2 \sin \frac{2\pi m}{k} \) &(\text{even } k), \\
\displaystyle
-\frac{\zeta(3)}{8\pi^2 k}+\frac{k}{4}\log 2-\frac{1}{k} \sum_{m=1}^{k-1} g_m(k)(k-g_m(k)) \log \(2 \sin \frac{\pi m}{k} \) \; &(\text{odd } k),
\end{cases}
\ea
\ee
where
\be
g_m(k)=\frac{k+(-1)^m(2m-k)}{4}.
\ee
Our conjecture \eqref{eq:A} predicts the small $N_f$ expansion
\be
A(N_f)=\frac{\zeta(3)}{\pi^2 N_f}-\frac{N_f}{24}+\(-\frac{\zeta(3)}{16\pi^2}+\frac{\log 2}{8} \)N_f^2
-\frac{\pi^2 N_f^3}{8640}+\cO(N_f^5).
\label{eq:A-WKB}
\ee
The leading term coincides with the result in \cite{GM}.
As we will see in section~\ref{sec:M2inst}, the next-to-leading correction is reproduced 
from the TBA analysis.
Also, in the large $N_f$ limit, we find
\be
\lim_{N_f \to \infty} \frac{A(N_f)}{N_f^2} = \frac{1}{8}\( -\frac{\zeta(3)}{\pi^2}+\log 2\)= 0.07141916904....
\ee
This result also agrees with the constant term in the genus-zero free energy found in \cite{GM}.
In figure~\ref{fig:A}, we plot $A(N_f)$ for finite $N_f$. 
The solid line is our conjecture, while
the dots are the numerical values of $A(N_f)$
extracted from the exact values of $Z(N_f,N)$, where
we estimated them by (see \eqref{eq:Zpert})
\be
A(N_f) \approx \log \left[ \frac{Z^\text{exact}(N_f,N)}{C(N_f)^{-1/3} \Ai \left[ C(N_f)^{-1/3}(N-B(N_f)) \right]} \right]\quad  (N \gg 1),
\label{eq:A-est}
\ee
for $N$ as large as possible.
For example, in the case of $N_f=3$, we computed exact $Z(3, N)$ up to $N=53$.
Using the estimation \eqref{eq:A-est} for $N=53$, we find
\be
A(3) \approx 0.60247027481429615744.
\ee
On the other hand, the exact value of $A(3)$ in \eqref{eq:A} is
\be
A(3)=-\frac{7 \zeta(3)}{12 \pi ^2}+\frac{1}{6} \log \frac{512}{9}.
\ee
The difference is $\cO(10^{-16})$, which is roughly the same order as the leading worldsheet instanton
correction $\cO(e^{-2\pi\sqrt{2N/N_f}})$.
For other values of $N_f$, our conjecture \eqref{eq:A} indeed show an agreement with 
the numerical estimation up to about $\cO(e^{-2\pi\sqrt{2N/N_f}})$.

\subsection{Non-perturbative corrections for integral $N_f$}\label{sec:Jnp-intNf}
Using the exact values of the partition function computed in subsection~\ref{sec:exact},
we can determine the non-perturbative correction to the grand potential.
The basic method is the same as that in \cite{HMO2}.
We first take an appropriate ansatz of the non-perturbative correction to the grand potential.
We then pull it back to the canonical partition function.
The coefficients in the ansatz are fixed by the numerical fitting of the exact values with high precision. 
See \cite{HMO2} in detail.

Since the partition function for $N_f=1,2$ is related to the ABJ(M) case, we can use the results in \cite{HMO2, HMO3, MM, HO}.
We also find the corrections explicitly for $N_f=3,4,6,8,12$, and numerically for other $N_f$'s.
We observe that all of these results take the form%
\footnote{One should not confuse this result with the worldsheet instanton correction \eqref{eq:JWS}.
In general, the coefficient $f_n(N_f,\mu)$ is the sum of all the contributions from the worldsheet instantons,
the membrane instantons and their bound states.}
\be
J_\text{np}(N_f,\mu)=\sum_{n=1}^\infty f_n(N_f,\mu) e^{-\frac{4n \mu}{N_f}},  \qquad N_f \in \mathbb{N} .
\ee
In particular, as in the ABJM case, the terms $e^{-(4m-2)\mu}$ ($m=1,2,\dots $) do not appear for odd $N_f$.
The explicit forms of $f_n(N_f,\mu)$ are complicated, and listed in appendix~\ref{sec:explicit}.

\subsection{Worldsheet instanton corrections}
Now let us consider the worldsheet instanton correction
\be
J^\text{WS}(N_f,\mu)=\sum_{m=1}^\infty f_{0,m}(N_f,\mu)e^{-\frac{4m\mu}{N_f}}.
\label{eq:JWS}
\ee
Here we give a conjecture of the analytic form of $f_{0,m}(N_f,\mu)$,
which is valid for general $N_f$, up to $m=3$.
To simplify the notation, we define
\be
s_n(N_f)=\sin\( \frac{2\pi n}{N_f} \),\qquad
P_n(N_f,\mu)=\frac{4n \mu+N_f}{\pi}.
\label{snPn}
\ee
Our conjecture of $f_{0,m}(N_f,\mu)$ up to $m=3$ is
\be
\ba
f_{0,1}&=-\frac{1}{2s_1}P_1,\\
f_{0,2}&=-\frac{1}{4}P_1^2+\frac{3s_3}{8 s_1 s_2} P_2-\frac{s_4}{2s_1^2 s_2},\\
f_{0,3}&= -\frac{s_2^2}{12 s_1}P_1^3+\frac{3s_4}{8 s_1} P_1 P_2
-\frac{5 s_4 s_5}{9 s_1 s_2 s_3}P_3-\frac{s_4^2}{2 s_1 s_2^2}P_1
-\frac{2s_6}{s_1^2 s_3}+\frac{2 s_4 s_5}{s_1^2 s_2^2}.
\ea
\label{eq:WSinst}
\ee
We have checked that this conjecture is consistent with the exact values of $Z(N_f,N)$
for $3\leq N_f\leq16$.
For $N_f=1,2$, worldsheet instantons \eqref{eq:WSinst} have poles already at the one-instanton level. 
We can also see that $f_{0,2}$ have poles at $N_f=1,2,4$
and $f_{0,3}$ have poles at $N_f=1,2,3,4,6.$\footnote{More precisely, $f_{0,2}$ (resp. $f_{0,3}$)
has poles at rational values of $N_f=1/n,2/n,4/n$ (resp. $N_f=1/n,2/n,3/n,4/n,6/n$) for all $n\in\mathbb{Z}$. Those poles should also be canceled by the higher membrane instantons.}
As we will see in the next section, those poles should be canceled by the membrane instantons.

\begin{figure}[tb]
\begin{center}
\begin{tabular}{cc}
\resizebox{70mm}{!}{\includegraphics{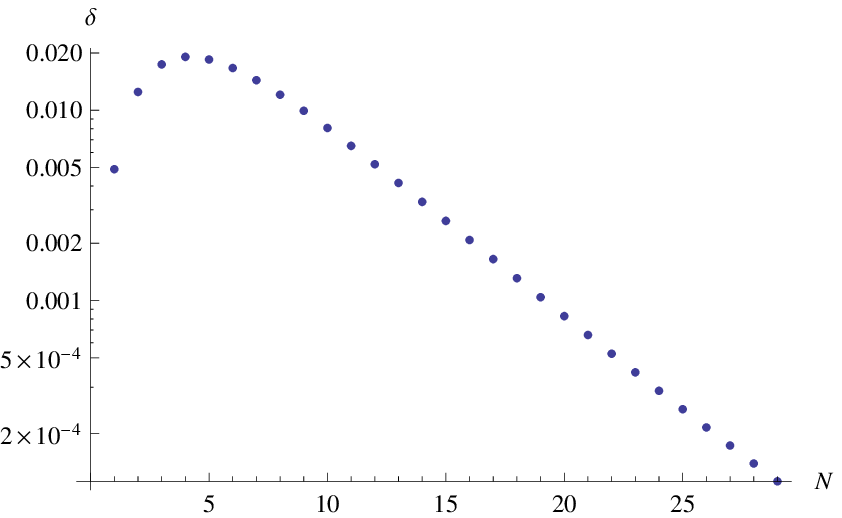}}
&
\resizebox{70mm}{!}{\includegraphics{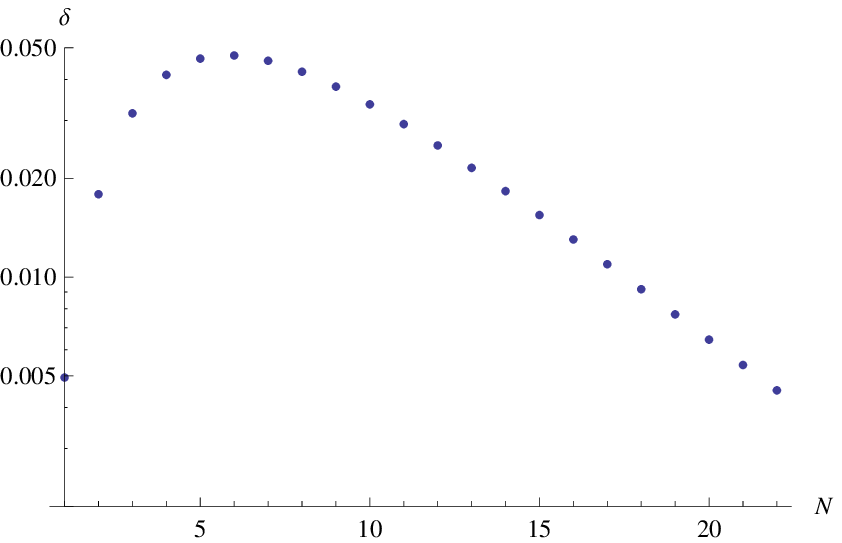}}
\\ {\footnotesize $N_f=7$} &  {\footnotesize $N_f=9$}
\vspace{7mm}
\\
\resizebox{70mm}{!}{\includegraphics{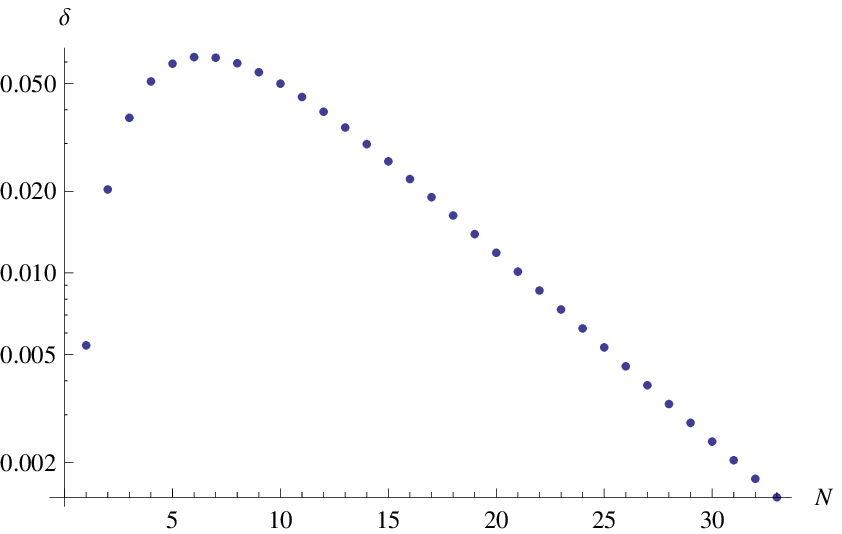}}
&
\resizebox{70mm}{!}{\includegraphics{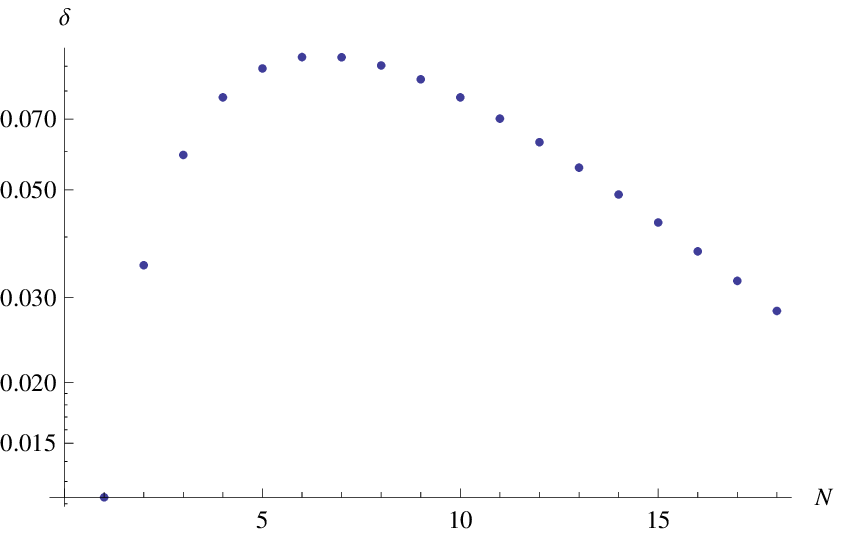}}
\\ {\footnotesize $N_f=10$} &  {\footnotesize $N_f=11$}
\vspace{7mm}
\\
\resizebox{70mm}{!}{\includegraphics{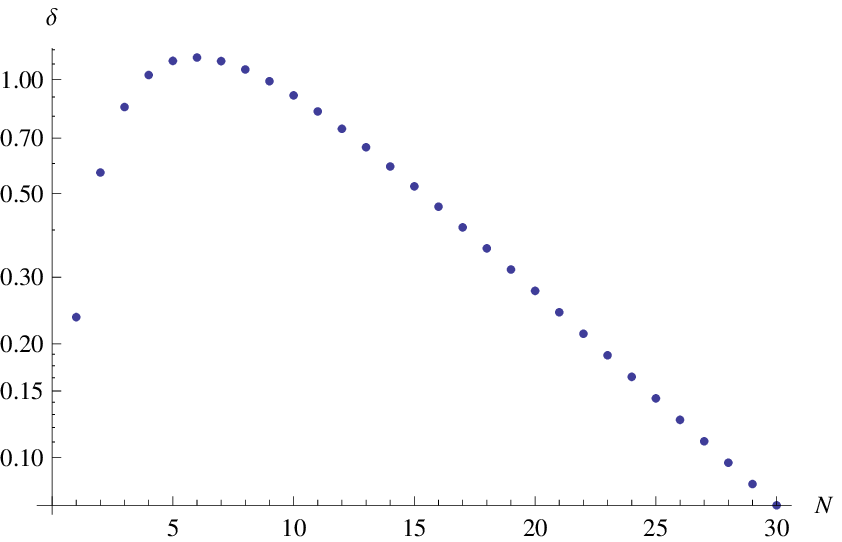}}
&
\resizebox{70mm}{!}{\includegraphics{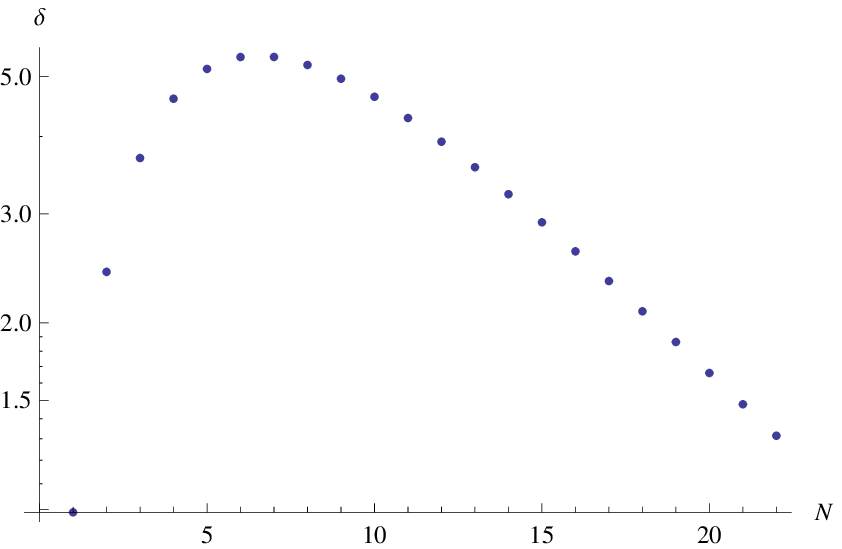}}
\\ {\footnotesize $N_f=14$} &  {\footnotesize $N_f=16$}
\end{tabular}
\end{center}
  \caption{We plot $\delta$ defined in \eqref{eq:delta} for $N_f=7,9,10,11,14,16$. 
Note that the vertical axis is log scale. One can see that $\delta$ is indeed
exponentially suppressed as $N$ grows.}
  \label{fig:delta}
\end{figure}

One can check that our conjecture \eqref{eq:WSinst}
correctly reproduces 
the result in appendix~\ref{sec:explicit}.
For instance, for $N_f=8$ we find
\be
\ba
f_{0,1}(8,\mu)&=-\frac{4\mu+8}{\rt{2}\pi},\quad
f_{0,2}(8,\mu)=-\frac{(4\mu+8)^2}{4\pi^2}+\frac{3(4\mu+4)}{4\pi}, \\
f_{0,3}(8,\mu)&=-\frac{(4\mu+8)^3}{6\rt{2}\pi^3}+4\rt{2},
\ea
\ee
which agree with the result \eqref{eq:Jnp-8} in appendix~\ref{sec:explicit}.
Other cases in appendix~\ref{sec:explicit}
are also reproduced by the conjecture \eqref{eq:WSinst}.%
\footnote{More precisely, the coefficient of $e^{-\alpha \mu}$ with $\alpha<2$ (even $N_f$) or $\alpha<4$ (odd $N_f$)
can be reproduced from \eqref{eq:WSinst}.
Beyond these values, we have to consider the membrane instanton and the bound state contributions.} 
We also stress that, for 
all other cases  $N_f=5,7,9,\cdots,16$,
our conjecture \eqref{eq:WSinst}
agrees highly non-trivially
with the instanton corrections extracted from the exact partition function.
To see it, let us define the non-perturbative correction to the partition function by
\be
Z(N_f,N)=Z_\text{pert}(N_f,N)(1+Z_\text{np}(N_f,N)),
\ee
where $Z_\text{pert}(N_f,N)$ is the perturbative contribution in \eqref{eq:Zpert},
neglecting all the exponentially suppressed corrections.
These non-perturbative corrections are encoded in $Z_\text{np}(N_f,N)$.
We have defined $Z_\text{np}(N_f,N)$ such that it decays exponentially in the large $N$ limit.
Note that for $N_f>6$, the first three corrections come from the worldsheet instantons
because $e^{-12\mu/N_f}>e^{-2\mu}$.
Namely,
\be
Z_\text{np}=Z_\text{WS}^{(1)}+Z_\text{WS}^{(2)}+Z_\text{WS}^{(3)}+(\text{subleading corrections}) 
\qquad (N_f>6),
\label{eq:Znp-3WS}
\ee
where $Z_\text{WS}^{(m)}$ is the worldsheet $m$-instanton correction.
As in the perturbative contribution, it is straightforward to translate 
the grand canonical result \eqref{eq:JWS} into the canonical one $Z_\text{WS}^{(m)}$.
We also introduce the quantity
\be
\delta=e^{6\pi \sqrt{2N/N_f}}\(\frac{Z}{Z_\text{pert}}-1-Z_\text{WS}^{(1)}-Z_\text{WS}^{(2)}
-Z_\text{WS}^{(3)}\).
\label{eq:delta}
\ee
Since the worldsheet $3$-instanton scales as the order $e^{-6\pi \sqrt{2N/N_f}}$, the subleading 
corrections in \eqref{eq:Znp-3WS} decay faster than $e^{-6\pi \sqrt{2N/N_f}}$.
Therefore the quantity $\delta$ must be exponentially suppressed in the large $N$ limit
for $N_f>6$ if our conjecture \eqref{eq:WSinst} is correct. 
In figure~\ref{fig:delta}, we plot $\delta$ for $N_f=7,9,10,11,14,16$ by using the exact data.
The quantity $\delta$ indeed decays exponentially when  $N$ is large, as expected.

As a further test, we consider the large $N_f$ limit of \eqref{eq:WSinst} with 
$\h{\mu}=\mu/N_f$ fixed.
It is not difficult to see
\be
\ba
f_{0,1}(N_f, \mu)&= \frac{N_f^2}{\pi^2} \(-\h{\mu}-\qu\)-\frac{2}{3}\h{\mu}-\frac{1}{6}+\cO(N_f^{-2}),\\
f_{0,2}(N_f, \mu)&= \frac{N_f^2}{\pi^2} \(-4\h{\mu}^2+\qu\h{\mu}-\frac{7}{32}\)-6\h{\mu}+\frac{11}{12}+\cO(N_f^{-2}) ,\\
f_{0,3}(N_f, \mu)&= \frac{N_f^2}{\pi^2} \(-\frac{128}{3}\h{\mu}^3+16\h{\mu}^2-\frac{46}{9}\h{\mu}+\frac{11}{27} \) \\
&\quad+\frac{1792 \h{\mu}^3}{9}-\frac{992 \h{\mu}^2}{3}+\frac{356 \h{\mu}}{3}-\frac{134}{9}+\cO(N_f^{-2}).
\ea
\label{eq:WS-'tHooft}
\ee
The leading terms exactly coincide with the genus zero contribution of the grand potential in the 't Hooft limit 
computed in appendix~\ref{sec:'tHooft}.
We stress that our conjecture \eqref{eq:WSinst} gives an all-genus prediction in the 't Hooft expansion.
It would be very interesting to confirm whether our conjecture indeed reproduces the higher genus corrections.\footnote{We are informed by A.~Grassi and M.~Mari\~{n}o that
they have computed the genus one free energy of this model.
Their result is consistent with our conjecture \eqref{eq:WSinst}.
We would like to thank them for sharing their unpublished result.}

From \eqref{eq:WSinst}, we expect that the worldsheet $m$-instanton coefficient $f_{0,m}(N_f,\mu)$
has the following general structure:
\begin{itemize}
 \item $f_{0,m}$ is an $m^\textrm{th}$ order polynomial of $\mu$, and the 
highest order term is $P_1^m$. 
\item  Most of the terms of $f_{0,m}$ have the same ``degree'' $m$, i.e. they have the form of
$\prod_lP_l^{k_l}$ with $\sum lk_l=m$, but some remaining terms have smaller degree
$\sum lk_l<m$.
\item The coefficient of each term in $f_{0,m}$ is a combination of $s_n~(n\leq2m)$.

\end{itemize}
It would be interesting to understand the origin of this structure and the
general rule to find the coefficients.

\section{Membrane instanton corrections}\label{sec:M2inst}
In the previous section, we extracted the non-perturbative corrections to the grand potential
from the exact values of the partition function for some integral values of $N_f$.
Based on these results, we proposed the analytic forms of the worldsheet instanton corrections
up to $m=3$.
As mentioned before, the grand potential also receives the non-perturbative corrections
from the membrane instantons.
In this section, we explore analytic forms of these corrections.
The membrane instanton correction corresponds to $m=0$ in \eqref{eq:Jnp},
\be
J^\text{M2}(N_f,\mu)=\sum_{\ell=1}^\infty f_{\ell,0}(N_f,\mu)e^{-2\ell \mu}.
\ee 
We want to determine the coefficient $f_{\ell,0}(N_f,\mu)$.
Unfortunately, we do not have a systematic way to compute $f_{\ell,0}(N_f,\mu)$.
Here, we try to fix it from many constraints.
The same idea was originally used in the ABJM Fermi-gas \cite{HMO2}.
We first investigate the expansion of the grand potential around $N_f=0$.
We then consider the singularity structure of $f_{\ell,0}(N_f,\mu)$.
Using these constraints, we present exact forms of $f_{1,0}(N_f,\mu)$
and a part of $f_{2,0}(N_f,\mu)$.
To fix the higher instanton corrections, we need more information.

\subsection{Semi-classical analysis from TBA}\label{sec:TBA-WKB}
We here study the expansion of the grand potential $J(N_f, z)$ around $N_f=0$.
Let us consider what kind of corrections the grand potential receives in the semi-classical limit $N_f \to 0$.%
\footnote{The term ``semi-classical limit'' is a bit confusing.
In the original coordinate, the commutation relation of the canonical variables $(x,p)$
is given by $[x,p]=2\pi i$, and thus $\hbar=2\pi$ is a constant in this coordinate.
However, as noted in \cite{Mezei:2013gqa,GM}, it is more convenient to rescale the position variable by $x=q/N_f$. 
In this new coordinate, the commutation relation becomes $[q,p]=2\pi i N_f$,
thus the limit $N_f \to 0$ corresponds to the semi-classical limit $\hbar=2\pi N_f \to 0$.
In the TBA, this rescale corresponds to the redefinition \eqref{eq:redef}.
} 
We first observe that $Z_1=\Tr \rho$ can be computed exactly \cite{GM},
\be
Z_1=\frac{1}{4\pi} \int_{-\infty}^\infty \frac{dx}{(2\cosh \frac{x}{2})^{N_f}}=\frac{1}{4\pi} \frac{\Gamma^2(N_f/2)}{\Gamma(N_f)}.
\ee
This has the following semi-classical expansion
\be
Z_1=\frac{1}{\pi N_f}-\frac{\pi N_f}{24}+\frac{\zeta(3)}{4\pi}N_f^2-\frac{\pi^3}{640}N_f^3
+\(-\frac{\pi \zeta(3)}{96}+\frac{3\zeta(5)}{16\pi} \)N_f^4+\cO(N_f^5).
\ee
This observation suggests that the grand potential also has the similar semi-classical expansion
\be
J(N_f, z)=\frac{1}{N_f}J_0(z)+N_f J_1(z)+N_f^2 J_{3/2}(z)+N_f^3 J_2(z)+N_f^4 J_{5/2}(z)+\cO(N_f^5),
\ee
because $J(N_f,z)$ is a kind of generating function of $Z_n=\Tr \rho^n$.
The absence of the constant $\cO(N_f^0)$ contribution is not obvious only from this observation, but the TBA analysis below
supports this.
The leading contribution $J_0(z)$ has already been computed in \cite{GM}.
Note that in the ABJM Fermi-gas, the even power terms $k^{2n}$ do not appear in the semi-classical
expansion.
This is a big difference between the $N_f$ matrix model and the ABJM case.
Here we compute $J_1(z)$ from the TBA equations.

What we should do is to solve the functional relations \eqref{eq:FR1} and \eqref{eq:FR2} in the semi-classical limit $N_f \to 0$
as in \cite{CM}.
To do so, we rescale all the functions by
\be
\ba
\widehat{R}_\pm(x)=R_\pm \( \frac{x}{N_f} \),\quad
\widehat{\eta}(x)=\eta \( \frac{x}{N_f} \),\quad
\widehat{U}(x)=U \( \frac{x}{N_f} \),
\ea
\label{eq:redef}
\ee
Then the equations \eqref{eq:FR1} and \eqref{eq:FR2} become
\be
\ba
\widehat{R}_+\( x+\frac{\pi i N_f}{2} \) \widehat{R}_+\(x- \frac{\pi i N_f}{2} \)
\exp \left[ \widehat{U}\( x+\frac{\pi i N_f}{2} \)+\widehat{U}\( x-\frac{\pi i N_f}{2} \) \right]
&=1+\widehat{\eta}^2(x),\\
\widehat{\eta}\( x+\frac{\pi i N_f}{2} \)+\widehat{\eta}\( x-\frac{\pi i N_f}{2} \)&=-z \widehat{R}_+(x),
\ea
\label{eq:FR3}
\ee
and
\be
\frac{\widehat{R}_-(x+\frac{\pi i N_f}{2} )}{\widehat{R}_+ ( x+\frac{\pi i N_f}{2} )}
-\frac{\widehat{R}_-(x-\frac{\pi i N_f}{2} )}{\widehat{R}_+ ( x-\frac{\pi i N_f}{2} )}
=2iN_f \frac{\widehat{\eta}'(x)}{1+\widehat{\eta}^2 (x)}.
\label{eq:FR4}
\ee
These are formally the same forms as the ones in (4.10) and (4.11) of \cite{CM} with $k = N_f$.
The only but big difference is the explicit form of the potential $\h{U}(x)$.

We assume that the functional equations \eqref{eq:FR3} and \eqref{eq:FR4} admit the semi-classical expansions
around $N_f=0$.
The important point is that the potential part in \eqref{eq:FR3} has the following ``semi-classical'' expansion
\be
\ba
\widehat{U}\( x+\frac{\pi i N_f}{2} \)+\widehat{U}\( x-\frac{\pi i N_f}{2} \)=|x|+N_f \log(1+e^{-\frac{2|x|}{N_f}} ) \qquad (N_f \to 0).
\ea
\label{eq:U-WKB}
\ee
The second term on the right hand side in \eqref{eq:U-WKB} looks like a ``non-perturbative'' term in $N_f$.
However, the integration of $e^{-2|x|/N_f}$ may potentially cause the perturbative corrections,
for example,
\be
\int_{-\infty}^\infty dx \, e^{-\frac{2|x|}{N_f}}=N_f.
\ee
Therefore we cannot drop the second term even in the semi-classical analysis.%
\footnote{This term makes the problem much harder than that in the ABJM case.
We will comment on the difficulty of the higher order computation later.}
As in \cite{CM}, we formally solve \eqref{eq:FR3} and \eqref{eq:FR4} by the semi-classical expansions,
\be
\ba
r(x)&=\widehat{R}_+(x)=\sum_{n=0}^\infty r_n(x) N_f^n, \qquad
\widehat{\eta}(x)=\sum_{n=0}^\infty \eta_n(x) N_f^n,\\
t(x)&=\frac{\widehat{R}_-(x)}{\widehat{R}_+(x)}=\sum_{n=0}^\infty t_n(x) N_f^n.
\ea
\label{eq:sol-WKB}
\ee
This can be done systematically up to any desired order in principle.
In appendix~\ref{sec:explicit}, we give explicit forms up to $n=3$.

Once the solutions of \eqref{eq:FR3} and \eqref{eq:FR4} are found,
we can compute the grand potential from \eqref{eq:Jpm-TBA}.
Let us define
\be
I_{\pm,n} \equiv \int_0^\infty \frac{dx}{2\pi} \widehat{R}_{\pm,n}(x),\qquad
\widehat{R}_{\pm}(x)=\sum_{n=0}^\infty N_f^n \widehat{R}_{\pm,n}(x),
\label{eq:Ipm}
\ee
where $\widehat{R}_{\pm,n}(x)$ are related to $r_n(x)$ and $t_n(x)$ in \eqref{eq:sol-WKB}.
Using the solutions in appendix~\ref{sec:explicit}, one can check that 
$I_{\pm, n}$ up to $n=3$ has the following expansions around $N_f=0$:
\be
\ba
I_{+,0}&=\frac{2}{\pi z} \arcsin \(\frac{z}{2} \),\qquad I_{-,0}=-\frac{2}{\pi^2 z} \arcsin^2 \(\frac{z}{2} \) ,\\
I_{\pm,1}&=N_f I_{\pm,1}^{(1)}+N_f^2 I_{\pm, 1}^{(2)}+\cO(N_f^3), \\
I_{\pm,2}&=I_{\pm, 2}^{(0)}+N_f I_{\pm, 2}^{(1)}+\cO(N_f^2),\\
I_{\pm,3}&=N_f^{-1} I_{\pm, 3}^{(-1)}+I_{\pm, 3}^{(0)}+\cO(N_f),
\ea
\label{eq:Ipm-WKB}
\ee
The explicit forms of these coefficients are also listed in appendix~\ref{sec:explicit}.
We also observe that the higher order corrections have the following expansions:
\be
\ba
I_{\pm, 2n}&=N_f^{-2n+3} I_{\pm, 2n}^{(-2n+3)}+\cO(N_f^{-2n+4}) \quad (n=2,3,\cdots), \\
I_{\pm, 2n+1}&=N_f^{-2n+2} I_{\pm, 2n+1}^{(-2n+2)}+\cO(N_f^{-2n+3}) \quad (n=2,3,\cdots).
\ea
\ee
Combining all the above results, the grand potential is given by
\be
\ba
J_z^{\pm}(z)&=\frac{1}{2\pi N_f} \int_0^\infty dx\, \widehat{R}_{\pm}(x)
=\frac{1}{N_f} \sum_{n=0}^\infty N_f^n I_{\pm, n} \\
&=\frac{I_{\pm,0}}{N_f}+N_f(I_{\pm,1}^{(1)}+I_{\pm,2}^{(0)}+I_{\pm, 3}^{(-1)})+N_f^2(I_{\pm,1}^{(2)}+I_{\pm,2}^{(1)}+I_{\pm,3}^{(0)}+\cdots)+\cO(N_f^3).
\ea
\ee
where $J_z^\pm(z)=\pd J^\pm (z)/\pd z$.

The leading term is given by
\be
J_{0,z}^+(z)=\frac{2}{\pi z} \arcsin \(\frac{z}{2} \),\qquad
J_{0,z}^-(z)=-\frac{2}{\pi^2 z} \arcsin^2 \(\frac{z}{2} \).
\ee
This result indeed agrees with (3.73) in \cite{GM}.
The next-to-leading term is also given by
\be
\ba
J_{1,z}^+(z)&=-\frac{\pi(16-z^2)}{12(4-z^2)^{5/2}}, \\
J_{1,z}^-(z)&=\frac{z(8-z^2)}{8(4-z^2)^2}+\frac{16-z^2}{6(4-z^2)^{5/2}}\arctan \( \frac{z}{\sqrt{4-z^2}} \),
\ea
\ee
After integrating over $z$, we finally obtain
\be
J_1(z)=\frac{z^2}{24(4-z^2)}-\frac{z(16-3z^2)}{48(4-z^2)^{3/2}} \left[\pi -2 \arctan\( \frac{z}{\sqrt{4-z^2}} \) \right], 
\ee
where we have fixed the integration constant such that $J_1(0)=0$.
Note that $J_1^+(z)$ and $J_1^-(z)$ have branch cuts for $|z|>2$.
However, the branch cut along $z>2$ disappears in the total potential $J_1(z)$,
and there is no discontinuity in $J_1(z)$ along the positive real axis in the $z$-plane.
In the large $z$ (or $\mu$) limit, $J_0(z)$ and $J_1(z)$ behave as
\be
\ba
J_0(z)&=\frac{2\mu^3}{3\pi^2}+\frac{\mu}{2}+\frac{\zeta(3)}{\pi^2}+\frac{2\mu+1}{\pi^2}e^{-2\mu}
+\frac{12\mu-1}{8\pi^2}e^{-4\mu}+\cO(\mu e^{-6\mu}),\\
J_1(z)&=-\frac{\mu}{8}-\frac{1}{24}-\frac{2\mu+1}{24} e^{-2\mu}+\frac{12\mu-19}{48} e^{-4\mu}+\cO(\mu e^{-6\mu}).
\ea
\label{eq:J0-largemu}
\ee
The cubic and linear terms in $\mu$ are consistent with the large $\mu$ behavior in \eqref{eq:J-largemu}.
The constant terms also match the expansion of $A(N_f)$ in \eqref{eq:A-WKB}.  
The exponentially suppressed terms will be used to fix the membrane instanton correction $f_{\ell,0}(N_f,\mu)$.

Let us remark on the higher order corrections.
The next-to-next-to-leading correction is given by
\be
J_{3/2,z}^\pm(z)=I_{\pm,1}^{(2)}+I_{\pm,2}^{(1)}+I_{\pm,3}^{(0)}+\cdots=\sum_{n=1}^\infty I_{\pm, n}^{(-n+3)}.
\ee
Thus to compute $J_{3/2}(z)$, we need the infinite series of corrections $I_{\pm, n}^{(-n+3)}$ ($n\geq 1$). 
This means that it is very difficult to compute the semi-classical expansion beyond this order
in this approach.
We need a more efficient way to resolve this problem.

One possible way is to expand the density matrix around $N_f=0$ from the beginning.
Let us see it briefly. We use the identity
\be
\ba
\frac{1}{(2\cosh \frac{x}{2N_f} )^{N_f/2}}=\exp \left[ -\frac{N_f}{2} \log \( 2\cosh \frac{x}{2N_f} \) \right] 
=e^{-\frac{|x|}{4}} \sum_{k=0}^\infty \frac{N_f^k}{k!} \ell^k(x),
\ea
\ee
where
\be
\ell(x) = -\frac{1}{2}\log (1+e^{-|x|/N_f}) .
\ee
We consider the rescaled density matrix
\be
\ba
\widetilde{\rho}(x_1,x_2)&=\frac{1}{N_f}\rho\( \frac{x_1}{N_f}, \frac{x_2}{N_f} \) \\
&=\frac{1}{2\pi N_f} \frac{1}{(2\cosh \frac{x_1}{2N_f} )^{N_f/2}} \frac{1}{(2\cosh \frac{x_2}{2N_f} )^{N_f/2}}
\frac{1}{2\cosh ( \frac{x_1-x_2}{2N_f} )}.
\ea
\ee
It is easy to check the equality $\Tr \widetilde{\rho}^n=\Tr \rho^n$.
Then, the density matrix is expanded as
\be
\ba
\widetilde{\rho}(x_1,x_2)
&=\frac{1}{2\pi N_f} \frac{e^{-\frac{|x_1|}{4}-\frac{|x_2|}{4}}}{2\cosh (\frac{x_1-x_2}{2N_f} )}
\sum_{k=0}^\infty \frac{N_f^k}{k!}(\ell(x_1)+\ell(x_2) )^k \\
&=\rho_0(x_1,x_2)+N_f \rho_1(x_1,x_2)+N_f^2 \rho_2(x_1,x_2)+\cdots,
\ea
\ee
where
\be
\rho_0(x_1,x_2)=\frac{1}{2\pi N_f} \frac{e^{-\frac{|x_1|}{4}-\frac{|x_2|}{4}}}{2\cosh (\frac{x_1-x_2}{2N_f} )},\qquad
\rho_k(x_1,x_2)=\frac{1}{k!}\rho_0(x_1,x_2) (\ell(x_1)+\ell(x_2) )^k.
\ee
The trace of $\widetilde{\rho}$ can be computed as follows:
\be
\ba
Z_n &= \Tr \widetilde{\rho}^n = \Tr [ (\rho_0+N_f \rho_1+N_f^2 \rho_2^2+\cdots)^n ] \\
&= \Tr \rho_0^n +n N_f \Tr \rho_0^{n-1} \rho_1+n N_f^2 \Tr \rho_0^{n-1} \rho_2 \\
&\quad+N_f^2 [ \Tr \rho_0^{n-2} \rho_1^2 + (\text{permutations}) ]+\cO(N_f^3).
\ea
\ee
Thus we find
\be
\ba
J(N_f,z)&=-\sum_{n=1}^\infty \frac{(-z)^n}{n} Z_n
= -\sum_{n=1}^\infty \frac{(-z)^n}{n} [\Tr \rho_0^n +n N_f \Tr \rho_0^{n-1} \rho_1+ \cdots ] \\
&= J_\text{ud}(N_f,z)-N_f \sum_{n=1}^\infty (-z)^n \Tr \rho_0^{n-1} \rho_1 +\cdots
\ea
\ee
where $J_\text{ud}(N_f,z)$ is the grand potential for the undeformed kernel $\rho_0$.
The second term is rewritten as
\be
\ba
\sum_{n=1}^\infty (-z)^n \Tr \rho_0^{n-1} \rho_1
&=2\sum_{n=1}^\infty (-z)^n \int_{-\infty}^\infty dx_1 \int_{-\infty}^\infty dx_2 \, \rho_0^{n-1}(x_1,x_2)
\rho_0(x_2,x_1) \ell(x_1) \\
&=2\sum_{n=1}^\infty (-z)^n \int_{-\infty}^\infty dx\, \rho_0^n(x,x) \ell(x)\\
&=-2 \int_{-\infty}^\infty dx  \frac{z \rho_0}{1+z \rho_0} (x,x) \ell(x).
\ea
\label{eq:Trrho01}
\ee
The semi-classical expansion of $\rho_0^n(x,x)$ can
be computed from the TBA for $\rho_0(x)$.

An advantage of this approach is that the undeformed kernel $\rho_0$ does not contain
the ``non-perturbative'' term in \eqref{eq:U-WKB}.
It is interesting to consider whether this approach resolves the difficulty of the computation of $J_{3/2}(z)$.

\subsection{Pole cancellation mechanism}
The worldsheet instanton correction conjectured in section~\ref{sec:np} has
singularities for some (in particular, integral) values of $N_f$.
These singularities must be canceled by the other contributions because
the partition function itself is always finite.
This pole cancellation mechanism was first found in the ABJM theory \cite{HMO2}.
As emphasized in \cite{CM}, this mechanism is conceptually important because
it implies that the 't Hooft expansion breaks down at some finite values of the string coupling $g_s$.
One needs to consider a non-perturbative completion to cure the divergences. 
Our conjecture \eqref{eq:WSinst} shows that this mechanism also exists in the 
$N_f$ matrix model.

A technical merit of this mechanism is that we can know the pole structure of the membrane
instanton correction from that of the worldsheet instanton correction.
As an example, let us consider the order $\cO(e^{-2\mu})$ correction for $N_f=2$.
Looking at \eqref{eq:Jnp}, there are two contributions at this order.
One is the worldsheet one-instanton correction $f_{0,1}$, and the other is
the membrane one-instanton correction $f_{1,0}$.
It is easy to see that the worldsheet one-instanton correction in \eqref{eq:WSinst} has the following
singularity at $N_f=2$.
\be
\lim_{N_f \to 2} f_{0,1}(N_f,\mu)e^{-\frac{4\mu}{N_f}}=-\frac{2(2\mu+1)}{\pi^2(N_f-2)}e^{-2\mu}-\frac{2(2\mu^2+2\mu+1)}{\pi^2}e^{-2\mu}+\cO(N_f-2).
\ee
This singularity must be canceled by the membrane one-instanton correction.
This means that $f_{1,0}$ must behave as
\be
\lim_{N_f \to 2} f_{1,0}(N_f,\mu)=\frac{2(2\mu+1)}{\pi^2(N_f-2)}+\cO(1).
\label{eq:f10-pole1}
\ee
Similarly, the singularities of $f_{1,0}$ at $N_f=4$ and $N_f=6$ are determined by the $f_{0,2}$ and $f_{0,3}$, respectively.
It is easy to find
\be
\ba
\lim_{N_f \to 4} f_{1,0}(N_f,\mu)&=\frac{6(2\mu+1)}{\pi^2(N_f-4)}+\cO(1),\\
\lim_{N_f \to 6} f_{1,0}(N_f,\mu)&=\frac{20(2\mu+1)}{\pi^2(N_f-6)}+\cO(1).
\ea
\label{eq:f10-pole2}
\ee
We will use these results to fix $f_{1,0}(N_f,\mu)$.

\subsection{Fixing the leading membrane instanton correction}
\begin{figure}[tb]
\begin{center}
\begin{tabular}{cc}
\resizebox{70mm}{!}{\includegraphics{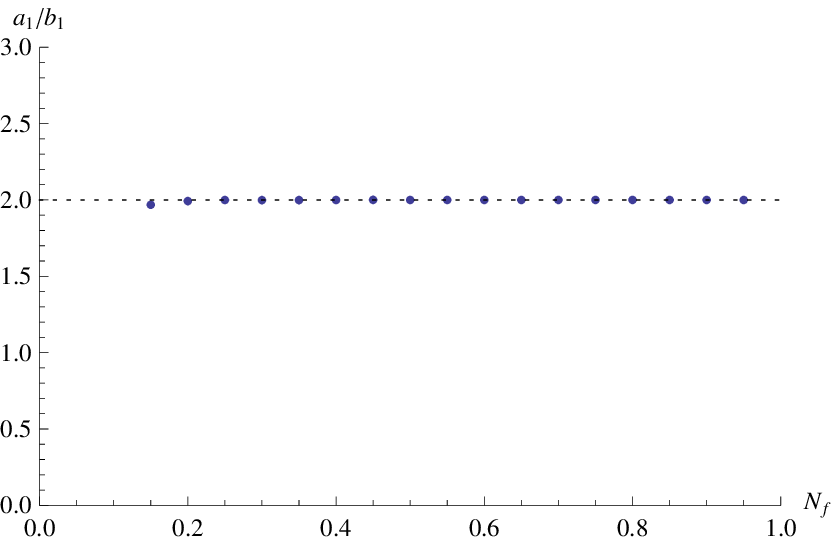}}
&
\resizebox{70mm}{!}{\includegraphics{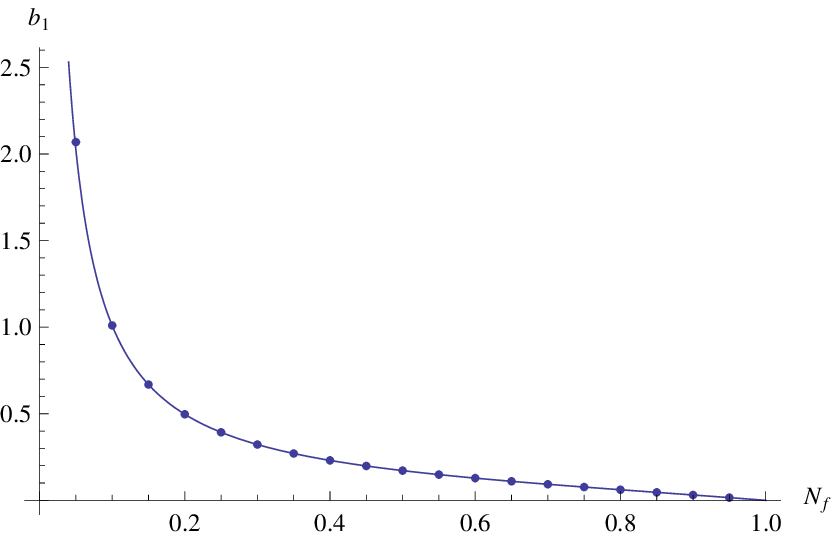}}
\vspace{3mm}
\\ (a) & (b)
\end{tabular}
\end{center}
  \caption{(a) We plot the ratio $a_1(N_f)/b_1(N_f)$ in \eqref{eq:f10-ansatz} against $N_f$. These values are estimated 
    by using the TBA equations \eqref{eq:TBA1} and \eqref{eq:TBA2}. The ratio does not depend on $N_f$.
   (b) We compare the analytic conjecture \eqref{eq:b1} with the numerically estimated values from TBA.}
  \label{fig:b1}
\end{figure}

Now we are in position to conjecture an exact form of $f_{1,0}(N_f,\mu)$.
The large $\mu$ expansion of the semi-classical results $J_0(z)$ and $J_1(z)$ 
suggests that the membrane $\ell$-instanton coefficient $f_{\ell,0}$ is a linear function of $\mu$
for all $\ell$
\be
f_{\ell,0}(N_f,\mu)=a_\ell(N_f)\mu+b_\ell(N_f).
\ee
To fix $f_{1,0}(N_f,\mu)$, we use the following three constraints:
\begin{itemize}
\item The singularity structure at $N_f=2,4,6$ is given by \eqref{eq:f10-pole1} and \eqref{eq:f10-pole2}.
\item As noted in subsection~\ref{sec:Jnp-intNf}, $f_{1,0}(N_f,\mu)$ must vanish for odd $N_f$.
\item The semi-classical expansion of $f_{1,0}(N_f,\mu)$ is given by
\be
f_{1,0}(N_f,\mu)=\frac{2\mu+1}{\pi^2N_f}-\frac{2\mu+1}{24}N_f+\cO(N_f^2) \qquad (N_f \to 0).
\label{eq:f10-WKB}
\ee
\end{itemize}
The results \eqref{eq:f10-pole1}, \eqref{eq:f10-pole2} and \eqref{eq:f10-WKB} strongly suggest that $f_{1,0}(N_f,\mu)$ has the following 
universal form for any $N_f$,
\be
f_{1,0}(N_f,\mu)=b_1(N_f)(2\mu+1),\qquad 
a_1(N_f)=2b_1(N_f).
\label{eq:f10-ansatz}
\ee
We can check that this is indeed the case.
In figure~\ref{fig:b1}(a), we show the ratio $a_1(N_f)/b_1(N_f)$ computed numerically from the TBA equations%
\footnote{We first compute $\Tr \rho^n$ numerically from the TBA equations as in \cite{HMO1}, 
and read off the numerical values
of the partition function. Then, we evaluate the values of the coefficients $a_1(N_f)$ and $b_1(N_f)$
in the same way in subsection~\ref{sec:Jnp-intNf}.}
 \eqref{eq:TBA1} and \eqref{eq:TBA2}
for $N_f=3/20,4/20,\dots,19/20$.
The ratio does not depend on $N_f$, and is very close to $2$.

Combining all the above results, we arrive at a conjecture of $b_1(N_f)$,
\be
b_1(N_f)=-\frac{\Gamma^2(-N_f/2)}{4\pi^2 \Gamma(-N_f)}.
\label{eq:b1}
\ee
The small $N_f$ expansion reads
\be
b_1(N_f)=\frac{1}{\pi ^2 N_f}-\frac{N_f}{24}-\frac{\zeta(3) N_f^2}{4 \pi ^2}
-\frac{\pi ^2 N_f^3}{640}+ \left(\frac{\zeta(3)}{96}-\frac{3 \zeta(5)}{16 \pi ^2}\right) N_f^4+\cO(N_f^5).
\ee
Of course, this reproduces \eqref{eq:f10-WKB}.
One non-trivial check of the conjecture \eqref{eq:b1} is to compute the finite parts for $N_f=2,4,6$.
For example, in the limit $N_f \to 2$, one finds
\be
\lim_{N_f \to 2} (f_{1,0}(N_f,\mu)e^{-2\mu}+f_{0,1}(N_f,\mu)e^{-\frac{4\mu}{N_f}})
=-\frac{4\mu^2+2\mu+1}{\pi^2}e^{-2\mu}.
\ee 
This result indeed reproduces the correct coefficient of $e^{-2\mu}$ in \eqref{eq:Jnp-2}.
In this way, one can also check that the coefficients of $e^{-2\mu}$ in \eqref{eq:Jnp-4} and \eqref{eq:Jnp-6}
are reproduced by the combinations $f_{1,0}e^{-2\mu}+f_{0,2}e^{-\frac{8\mu}{N_f}}$ ($N_f \to 4$) and
$f_{1,0}e^{-2\mu}+f_{0,3}e^{-\frac{12\mu}{N_f}}$ ($N_f \to 6$), respectively.
As another check, we compare the numerical values computed from TBA
with our conjecture \eqref{eq:b1} for various values of $N_f$.
The result is shown in figure~\ref{fig:b1}(b).
These tests present strong supports for our conjecture \eqref{eq:b1}.

\subsection{Remark on higher instanton corrections}
\begin{figure}[tb]
\begin{center}
\begin{tabular}{cc}
\resizebox{70mm}{!}{\includegraphics{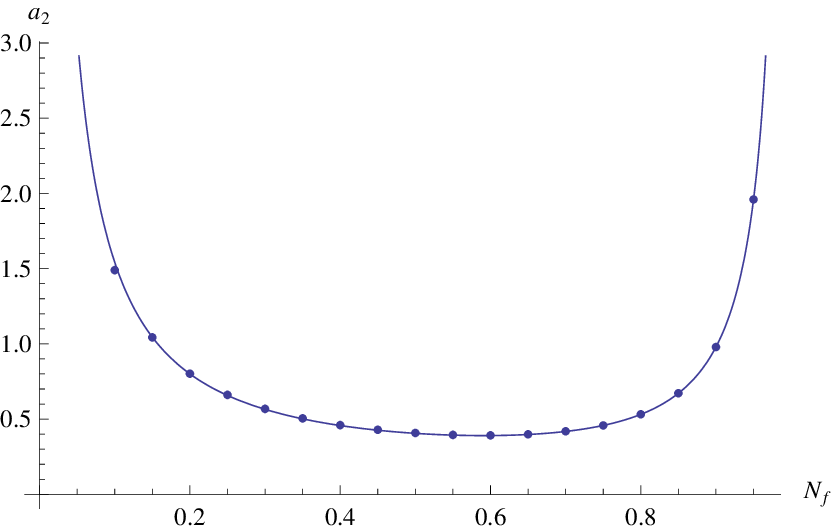}}
&
\resizebox{70mm}{!}{\includegraphics{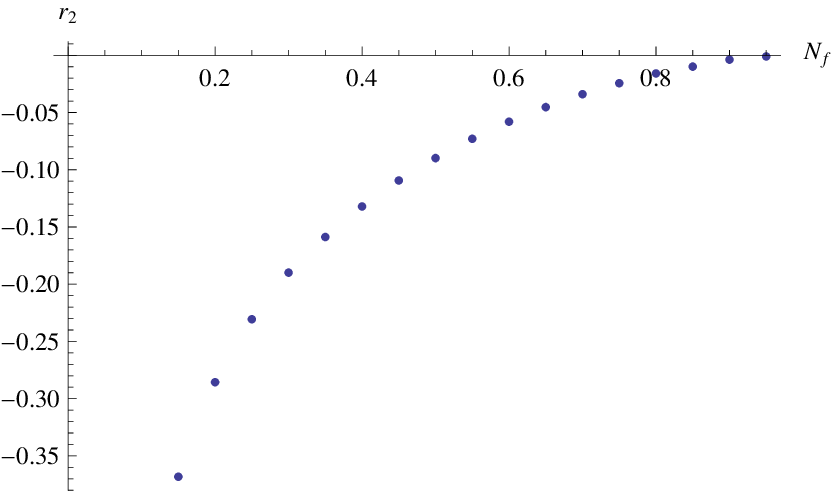}}
\vspace{5mm}
\\ (a) & (b)
\end{tabular}
\end{center}
  \caption{We plot (a) $a_2(N_f)$ and (b) $r_2(N_f)=b_2(N_f)-a_2(N_f)/4$ against $N_f$.}
  \label{fig:a2}
\end{figure}

Let us remark on the higher instanton corrections.
So far, we do not know how to determine $a_\ell(N_f)$ and $b_\ell(N_f)$ systematically.
Let us consider the next simplest case: $\ell=2$.
From \eqref{eq:J0-largemu}, these must have the small $N_f$ expansions,
\be
\ba
a_2(N_f)=\frac{3}{2\pi^2 N_f}+\frac{1}{4}N_f+\cO(N_f^2) ,\qquad
b_2(N_f)=-\frac{1}{8\pi^2 N_f}-\frac{19}{48}N_f+\cO(N_f^2).
\ea
\ee
Furthermore, $f_{2,0}(N_f,\mu)$ must cancel the poles of $f_{0,1}(N_f,\mu)$ at $N_f=1$ and $f_{0,3}(N_f,\mu)$ at $N_f=3$.
These conditions give the constraints
\be
\ba
\lim_{N_f \to 1}a_2(N_f)&=-\frac{1}{\pi^2(N_f-1)}+\cO(1),\qquad 
\lim_{N_f \to 1}b_2(N_f)=-\frac{1}{4\pi^2(N_f-1)}+\cO(1), \\
\lim_{N_f \to 3}a_2(N_f)&=-\frac{10}{\pi^2(N_f-3)}+\cO(1),\qquad
\lim_{N_f \to 3}b_2(N_f)=-\frac{5}{2\pi^2(N_f-3)}+\cO(1).
\ea
\label{eq:poles-2M2}
\ee
We find an analytic form of $a_2(N_f)$ satisfying all these constraints
\be
a_2(N_f)=-\frac{1}{4\pi^2}\(1+\frac{2}{\cos \pi N_f} \) \frac{\Gamma^2(-N_f)}{\Gamma(-2N_f)}.
\label{eq:a2}
\ee
This guess correctly reproduces the finite parts for $N_f=1,3$ in appendix~\ref{sec:explicit}.
It also shows a good agreement with the numerical values from TBA as shown in figure~\ref{fig:a2}(a).

We also want to fix $b_2(N_f)$. 
Unfortunately, the ratio $b_2(N_f)/a_2(N_f)$ is no longer a constant unlike the one-instanton correction.
Instead, it is convenient to define a new function by
\be
r_\ell(N_f)=b_\ell(N_f)-\frac{a_\ell(N_f)}{2\ell}.
\ee
In other words, we parametrize the 
membrane $\ell$-instanton coefficient as
\be
f_{\ell,0}(N_f,\mu)=\Big(\mu+\frac{1}{2\ell}\Big)a_\ell(N_f)+r_\ell(N_f).
\ee
For $\ell=1$,
we found in \eqref{eq:f10-ansatz} that the last term is absent:
$r_1(N_f)=0$.
For $\ell\geq2$,
it turns out that
$r_\ell(N_f)$ is a non-trivial function of $N_f$.
From the pole structure \eqref{eq:poles-2M2}, one finds that the function $r_2(N_f)$ is regular
at $N_f=1,3$.
Furthermore, in order to reproduce the finite parts for $N_f=1,3$ in \eqref{eq:Jnp-1} and \eqref{eq:Jnp-3} correctly,
$r_2(N_f)$ must vanish in the limit $N_f\to 1,3$:
\be
\lim_{N_f \to 1}r_2(N_f)= \lim_{N_f \to 3}r_2(N_f)=0.
\ee
Also, $r_2(N_f)$ must have the following semi-classical expansion
\be
r_2(N_f)=-\frac{1}{2\pi^2 N_f}-\frac{11}{24}N_f+\cO(N_f^2) \qquad (N_f \to 0).
\ee 
In figure~\ref{fig:a2}(b), we plot $r_2(N_f)$ for $0<N_f<1$ as a function of $N_f$.
As expected, $r_2(N_f)$ goes to zero as $N_f \to 1$.
So far, we have not found an exact form of $r_2(N_f)$.
To fix it, we might need more information.

Interestingly, the combination $(\mu+\frac{1}{2n})a_n(N_f)$ seems to cancel the poles of the worldsheet instantons
when there are no bound state contributions.
For example, for $N_f=2/n$ ($n=1,2,3,\dots$), the worldsheet 1-instanton correction has the following pole structure:
\be
\ba
\lim_{N_f \to 2/n} f_{0,1}(N_f,\mu)e^{-\frac{4\mu}{N_f}}=(-1)^{n}\frac{4(\mu+\frac{1}{2n})}{n^2\pi^2(N_f-\frac{2}{n})}e^{-2n \mu}+\cO(1).
\ea
\label{eq:pole-1WS-frac}
\ee
This pole must be canceled by the membrane $n$-instanton correction because there are no bound state contributions
at this order.
Thus, focusing on the $\mu e^{-2n \mu}$ term, one finds that $a_n(N_f)$ must behave at $N_f=2/n$ as
\be
\ba
\lim_{N_f \to 2/n}a_n(N_f)=(-1)^{n-1}\frac{4}{n^2\pi^2(N_f-\frac{2}{n})}+\cO(1),
\ea
\ee
and that the combination $(\mu+\frac{1}{2n})a_n(N_f)$ indeed cancels the pole in \eqref{eq:pole-1WS-frac}.
Therefore we conclude that $r_n(N_f)$ is regular at $N_f=2/n$:
\be
\lim_{N_f \to 2/n}r_n(N_f)=\cO(1).
\ee
It would be interesting to clarify the analytic structure of $r_n(N_f)$ in more detail.

\section{Conclusions}
In this paper we have studied the large $N$ non-perturbative effects
in the $N_f$ matrix model.
Combining the exact computation of the partition function and the analysis of TBA,
we have successfully determined a first few
terms of both the 
worldsheet instanton corrections and the membrane instanton corrections, 
as analytic functions of $N_f$.
These analytic results show that the pole cancellation mechanism, originally found in the ABJM model \cite{HMO2}, works also for the $N_f$ matrix model.
As emphasized in \cite{GM},
this clearly shows that the 't Hooft expansion alone is incomplete
and the existence of both membrane instantons and worldsheet instantons
is necessary for the non-perturbative definition of the theory and
the free energy to be finite at the physical value of the coupling.
We would like to emphasize that this mechanism is invisible
as long as one focuses on the small/large $N_f$ limit.
One needs to consider the M-theory (or finite coupling) regime to see it.
It is desirable to 
check that this mechanism works more generally,
by computing 
the higher instanton coefficients of worldsheet instantons, membrane instantons, 
and their bound states.
It is also interesting to apply the Fermi-gas approach to BPS Wilson loops
along the lines of \cite{Klemm:2012ii, Hatsuda:2013yua, Hirano:2014bia}.
Moreover, it would be interesting to explore the non-perturbative structure for other examples \cite{Mezei:2013gqa, Anderson:2014hxa}.

To proceed further, we need to develop a systematic way to study this model.
In particular, it is very interesting to find an efficient method to 
compute the small $N_f$ expansion of the grand potential, which gives important 
clues to determine the analytic form of the membrane instantons.
As for the worldsheet instantons, using the technique of ordinary matrix models
we can in principle compute the genus expansion in the 't Hooft limit, order by order
in $1/N_f$. However, to discuss the pole cancellation,
we need to resum this series {\it \`{a} la} Gopakumar-Vafa, which is a formidable task.
Also, currently we do not have enough information to 
study the effect of bound states of worldsheet instantons and  membrane instantons. 
We leave them as interesting future problems.

We found that the structure of the instanton corrections
in the $N_f$ matrix model is quite different from that of the ABJM model.
In the ABJM case, the membrane instanton correction and the worldsheet instanton correction have the following form \cite{MP2, HMO2}
\be
\ba
J^\text{M2}_\text{ABJM}(k,\mu)&=\sum_{\ell=1}^\infty (a_\ell(k)\mu^2+b_\ell(k) \mu+c_\ell(k))e^{-2\ell \mu},\\
J^\text{WS}_\text{ABJM}(k,\mu)&=\sum_{m=1}^\infty d_m(k) e^{-\frac{4m\mu}{k}},
\ea
\ee
where all the coefficients $a_\ell(k)$, $b_\ell(k)$, $c_\ell(k)$ and $d_m(k)$ are expressed in terms of a combination of the trigonometric
functions.
On the other hand, 
in the case of $N_f$ matrix model, we conjecture that 
the membrane instanton correction and the worldsheet instanton correction have
the following structure
\be
\ba
 J^\text{M2}(N_f,\mu)&=\sum_{\ell=1}^\infty (a_\ell(N_f)\mu+b_\ell(N_f) )e^{-2\ell \mu},\\
J^\text{WS}(N_f,\mu)&=\sum_{m=1}^\infty \Big(\sum_{\sum lk_l\leq m} d_m^{(k_1,\cdots,k_m)}(N_f)
\prod_{l=1}^m P_l^{k_l}(N_f,\mu)\Big)e^{-\frac{4m\mu}{N_f}},
\label{Jconjecture}
\ea
\ee
where $P_l(N_f,\mu)$ is defined in \eqref{snPn}. 
We found that the membrane instanton coefficients
in \eqref{eq:b1} and \eqref{eq:a2} are expressed in terms of the gamma function 
(and the trigonometric functions). As another difference, the pre-factor of the worldsheet instanton 
correction depends on $\mu$, unlike the ABJM case. More precisely, the worldsheet $m$-instanton coefficient
is given by an $m^\text{th}$ order polynomial of $\mu$, and the coefficient
$ d_m^{(k_1,\cdots,k_m)}(N_f)$ appearing
in \eqref{Jconjecture} is a combination of the trigonometric functions \eqref{eq:WSinst}.
This difference between the $N_f$ matrix model and the ABJM model
may be related to the difference of the orbifolding on the bulk side:
the orbifold $\mathbb{C}^2\times \mathbb{C}^2/\mathbb{Z}_{N_f}$, 
corresponding to the $N_f$ matrix model,
has a family of $A_{N_f-1}$ ALE singularity parametrized by the first factor $\mathbb{C}^2$, 
while the singularity of $\mathbb{C}^4/\mathbb{Z}_k$ for the ABJM case
is isolated. Perhaps, the structure of the worldsheet instantons 
in the $N_f$ matrix model might be understood as the effect of a non-isolated family
of rational curves (see \cite{Beasley:2005iu} for such worldsheet instanton effects
in a heterotic string compactification).

The spectral problem in the Fermi-gas is also important.
The spectrum of the one-dimensional Fermi-gas that we are considering is determined by
the Fredholm integral equation of the first kind \cite{MP2}:
\be
\int_{-\infty}^\infty dx' \rho(x,x') \phi(x')=e^{-E} \phi(x).
\label{eq:int-eq-spec}
\ee
Since the density matrix $\rho(x_1,x_2)$ is a non-negative Hilbert-Schmidt operator,
the integral equation \eqref{eq:int-eq-spec} has a positive discrete spectrum.
In the ABJ(M) Fermi-gas, the spectrum is determined by the exact WKB quantization 
condition \cite{KM, Kallen:2014lsa} (see also \cite{Huang:2014eha}),
in which one has to consider not only the perturbative contribution but also the non-perturbative contribution in the Planck constant $\hbar=2\pi k$ to reproduce the correct spectrum. 
Since all the information of the (grand) partition function is encoded in the Fermi-gas spectrum,
it is important to find the exact WKB quantization 
condition in the $N_f$ matrix model.

In the case of ABJ(M) model, 
we have a very detailed understanding of the
instanton corrections thanks to the relation to the refined topological string on
local $\mathbb{P}^1\times \mathbb{P}^1$ 
\cite{HMMO,MM,HO,KM,Kallen:2014lsa}.
This relation to the topological string is widely viewed as an accident of the ABJ(M) model.
However, in view of the non-trivial relation \eqref{eq:A}
between the constant term $A(N_f)$ of the $N_f$ matrix model and the constant map contribution 
$A_{\rm const}(k)$
of the topological string,
it is tempting to speculate that the $N_f$ matrix model also has a hidden connection to the topological string
on some background.
It would be interesting to see if
such a hidden connection to the topological string really exists, or not.

Finally, in this paper, we have probed the non-perturbative effects in M-theory from
its gauge theory dual.
Recently, there are some interesting progress on the gravity side \cite{Bhattacharyya:2012ye, Dabholkar:2014wpa}.
It would be very significant to confirm our pole cancellation mechanism directly
in M-theory in the future.

\acknowledgments{We would like to thank Alba Grassi and Marcos Mari\~{n}o for correspondence.
We are grateful to Marcos Mari\~{n}o for helpful discussions and comments on the manuscript.
We are also grateful to Satoru Odake for allowing us to use computers in
the theory group, Shinshu University.
The work of K.O. is  supported in part by JSPS Grant-in-Aid for Young Scientists
(B) 23740178.
}

\appendix

\section{A simpler expression of the constant map}\label{sec:Aconst}

In this appendix, we derive the integral expression \eqref{eq:A-int} of the constant map $A_\text{const}(k)$
in the ABJM Fermi-gas (or equivalently in the topological string on local $\mathbb{P}^1 \times \mathbb{P}^1$).
A similar integral expression was found in \cite{KEK}, but our expression is much simpler than theirs. 
Our starting point is the all-order small $k$ expansion found in \cite{KEK},
\be
A_\text{const}(k)=\frac{2\zeta(3)}{\pi^2 k}+\sum_{n=1}^\infty \frac{(-1)^n}{(2n)!}B_{2n}B_{2n-2}\pi^{2n-2}k^{2n-1}.
\ee
To rewrite this as an integral form, we use the identity for the Bernoulli number:
\be
B_{2n}=(-1)^{n-1}4n \int_0^\infty dx \frac{x^{2n-1}}{e^{2\pi x}-1} \quad (n\geq 1).
\ee
This identity is simply obtained from the integral expression of $\zeta(z)$ by setting $z=2n$.
Then, we get
\be
\ba
\sum_{n=1}^\infty \frac{(-1)^n}{(2n)!}B_{2n}B_{2n-2}\pi^{2n-2}k^{2n-1}
= -\frac{2}{\pi} \int_0^\infty dx \frac{1}{e^{2\pi x}-1} \sum_{n=1}^\infty \frac{B_{2n-2}}{(2n-1)!}(\pi k x)^{2n-1}.
\ea
\ee
The sum can be performed,
\be
\sum_{n=1}^\infty \frac{B_{2n-2}}{(2n-1)!} z^{2n-1}=\frac{z^2}{4}+\Li_2(1-e^{-z}).
\ee
Therefore we obtain the integral form
\be
\ba
A_\text{const}(k)=\frac{2\zeta(3)}{\pi^2 k}\(1-\frac{k^3}{16}\) -\frac{2}{\pi} \int_0^\infty dx \frac{\Li_2(1-e^{-\pi k x})}{e^{2\pi x}-1}.
\ea
\ee
After integration by parts, we finally get \eqref{eq:A-int}.

\section{Some explicit results}\label{sec:explicit}
\subsection{Corrections for integral $N_f$}
Here we summarize the explicit forms of $J_\text{np}(N_f,\mu)$ for $N_f=1,2,3,4,6,8,12$.
For $N_f=1$, the partition function is equivalent to the one in the ABJM theory at $k=1$.
We can use the result in \cite{HMO3},%
\footnote{There is a typo in version 2 of \cite{HMO3}. The term $-292064/3$ in the coefficient
of $e^{-24\mu}$ must be $+292064/3$.}
\begin{align}
J_\text{np}(1,\mu)&=\biggl[\frac{4\mu^2+\mu+1/4}{\pi^2}\biggr]e^{-4\mu}
+\biggl[-\frac{52\mu^2+\mu/2+9/16}{2\pi^2}+2\biggr]e^{-8\mu}\nn
&\quad+\biggl[\frac{736\mu^2-152\mu/3+77/18}{3\pi^2}-32\biggr]e^{-12\mu}\nn
&\quad+\biggl[-\frac{2701\mu^2-13949\mu/48+11291/768}{\pi^2}+466\biggr]e^{-16\mu} \label{eq:Jnp-1}\\
&\quad+\biggl[\frac{161824\mu^2-317122\mu/15+285253/300}{5\pi^2}-6720\biggr]e^{-20\mu} \nn
&\quad+\biggl[-\frac{1227440\mu^2-2686522\mu/15+631257/80}{3\pi^2}+\frac{292064}{3}\biggr]e^{-24\mu} 
+\cO(e^{-28\mu}). \nonumber
\end{align}
Similarly, in the case of $N_f=2$, we can use the ABJ result in \cite{MM},\footnote{%
For $N_f=1,2$, one can check that
the coefficients of $J_{\rm np}(\mu)$ are reproduced from the general formula in \cite{HO} by plugging in the explicit values of 
the refined BPS invariants $N_{j_L,j_R}^{d_1,d_2}$ of local $\mathbb{P}^1\times\mathbb{P}^1$.}
\begin{align}
J_\text{np}(2,\mu)&=\biggl[-\frac{4\mu^2+2\mu+1}{\pi^2}\biggr]e^{-2\mu}
+\biggl[-\frac{52\mu^2+\mu+9/4}{2\pi^2}+2\biggr]e^{-4\mu}\nn
&\quad+\biggl[-\frac{736\mu^2-304\mu/3+154/9}{3\pi^2}+32\biggr]e^{-6\mu} \nn
&\quad+\biggl[-\frac{2701\mu^2-13949\mu/24+11291/192}{\pi^2}+466\biggr]e^{-8\mu}\label{eq:Jnp-2}\\
&\quad+\biggl[-\frac{161824\mu^2-634244\mu/15+285253/75}{5\pi^2}+6720\biggr]e^{-10\mu} \nn
&\quad+\biggl[-\frac{1227440\mu^2-5373044\mu/15+631257/20}{3\pi^2}+\frac{292064}{3}\biggr]e^{-12\mu} 
+\cO(e^{-14\mu}). \nonumber
\end{align}
For $N_f=3,4,6,8,12$, we find new results
\begin{align}
J_\text{np}(3,\mu)&=-\frac{4\mu+3}{\rt{3}\pi}e^{-4\mu/3}+
\left[-\frac{(4\mu+3)^2}{4\pi^2}+\frac{2}{3}\right]e^{-8\mu/3}\nn
&\quad+\left[-\frac{(4\mu+3)^3}{8\rt{3}\pi^3}+\frac{76\mu^2}{3\pi^2}+\frac{47(4\mu+1)}{24\pi^2}
-\frac{4\mu+3}{\rt{3}\pi}-8\right]e^{-4\mu}\nn
&\quad+\left[-\frac{(4\mu+3)^4}{32\pi^4}-\frac{(4\mu+3)^2}{2\pi^2}
+\frac{166\mu+133/8}{\rt{3}\pi}+\frac{2}{3}\right]e^{-16\mu/3}\label{eq:Jnp-3}\\
&\quad+\biggl[-\frac{\rt{3}(4\mu+3)^5}{160\pi^5}-\frac{(4\mu+3)^3}{\rt{3}\pi^3}
+\frac{332\mu^2+1129\mu/4+399/16}{\pi^2} \nn
&\qquad\quad+\frac{6(4\mu+3)}{\rt{3}\pi}-48\biggr]e^{-20\mu/3}+\cO(e^{-8\mu}), \nn
J_\text{np}(4,\mu)&=-\frac{4\mu+4}{2\pi}e^{-\mu}+\left[-\frac{10\mu^2+7\mu+7/2}{\pi^2}+1\right]e^{-2\mu}
-\frac{88\mu+52/3}{3\pi}e^{-3\mu}\label{eq:Jnp-4}\\
&\quad+\left[-\frac{269\mu^2+193\mu/4+265/16}{\pi^2}+58\right]e^{-4\mu}
-\frac{4792\mu+1102/5}{5\pi}e^{-5\mu}+\cO(e^{-6\mu}),\nn
J_\text{np}(6,\mu)&=-\frac{4\mu+6}{\rt{3}\pi}e^{-2\mu/3}
+\left[-\frac{(4\mu+6)^2}{4\pi^2}+\frac{2}{3}\right]e^{-4\mu/3}\nn
&\quad+\left[-\frac{(4\mu+6)^3}{8\rt{3}\pi^3}-
\frac{76\mu^2}{3\pi^2}-\frac{47(2\mu+1)}{6\pi^2}
-\frac{4\mu+6}{\rt{3}\pi}+8\right]e^{-2\mu}\nn
&\quad+\left[-\frac{(4\mu+6)^4}{32\pi^4}-\frac{(4\mu+6)^2}{2\pi^2}
-\frac{166\mu+133/4}{\rt{3}\pi}+\frac{2}{3}\right]e^{-8\mu/3}\label{eq:Jnp-6}\\
&\quad+\biggl[-\frac{\rt{3}(4\mu+6)^5}{160\pi^5}-\frac{(4\mu+6)^3}{\rt{3}\pi^3}
-\frac{332\mu^2+1129\mu/2+399/4}{\pi^2}\nn
&\qquad \quad+\frac{6(4\mu+6)}{\rt{3}\pi}+48\biggr]e^{-10\mu/3}+\cO(e^{-4\mu}),\nn
J_\text{np}(8,\mu)&=-\frac{4\mu+8}{\rt{2}\pi}e^{-\mu/2}+\left[-\frac{(4\mu+8)^2}{4\pi^2}
+\frac{3(4\mu+4)}{4\pi}\right]e^{-\mu}\label{eq:Jnp-8} \\
&\quad+\left[-\frac{(4\mu+8)^3}{6\rt{2}\pi^3}+4\rt{2}\right]e^{-3\mu/2}+\cO(e^{-2\mu}), \nn
J_\text{np}(12,\mu)&=-\frac{4\mu+12}{\pi}e^{-\mu/3}
+\left[-\frac{(4\mu+12)^2}{4\pi^2}
+\frac{\rt{3}(8\mu+12)}{2\pi}-2\right]e^{-2\mu/3}\label{eq:Jnp-12}\\
&+\left[-\frac{(4\mu+12)^3}{8\pi^3}+\frac{3\rt{3}(4\mu+12)(8\mu+12)}{8\pi^2}
-\frac{32\mu+56}{3\pi}+\frac{8}{\rt{3}}\right]e^{-\mu}+\cO(e^{-4\mu/3}). \nonumber
\end{align}

\subsection{Semi-classical solutions of TBA}\label{sec:TBA-WKB-sol}
Here we give the solutions \eqref{eq:sol-WKB} of the TBA equations in the semi-classical limit.
For simplicity, we introduce
\be
L(x)=\log(1+e^{-\frac{2|x|}{N_f}}),
\ee
Note that the solutions below are valid for $x>0$.
Since the solutions are invariant under $x \to -x$,
it is easy to know the solutions for $x<0$.
The solutions up to $n=3$ are given by 
\begin{align}
r_0(x)&=\frac{2}{\sqrt{4e^x-z^2}},\quad \eta_0(x)=-\frac{z}{\sqrt{4e^x-z^2}},\quad
t_0(x)=-\frac{2}{\pi} \arctan \( \frac{z}{\sqrt{4e^x-z^2}} \), \\
r_1(x)&=-\frac{4 e^x L(x)}{\left(4 e^x-z^2\right)^{3/2}},\quad
\eta_1(x)=\frac{2 e^x z L(x)}{\left(4 e^x-z^2\right)^{3/2}},\quad
t_1(x)=\frac{z L(x)}{\pi \sqrt{4e^x-z^2}},\\
r_2(x)&=\frac{\pi^2 z^2 e^x(6e^x+z^2)}{2(4e^x-z^2)^{7/2}}+\frac{2e^x(2e^x+z^2)}{(4e^x-z^2)^{5/2}}L^2(x) ,\nn
\eta_2(x)&=-\frac{2\pi^2 z e^{2x}(e^x+z^2)}{(4e^x-z^2)^{7/2}}-\frac{ze^x (2e^x+z^2)}{(4e^x-z^2)^{5/2}}L^2(x) ,\\
t_2(x)&=-\frac{\pi z e^x(16e^x+11z^2)}{12(4e^x-z^2)^{5/2}}-\frac{z e^x}{\pi (4e^x-z^2)^{3/2}}L^2(x),\nn
r_3(x)&=-\frac{\pi ^2 z^2 e^x (36 e^{2 x}+22 e^x z^2+z^4)}{2 (4 e^x-z^2)^{9/2}}L(x)-\frac{2 e^x (4 e^{2 x}+10 e^x z^2+z^4)}{3(4 e^x-z^2)^{7/2}}L^3(x)\nn
&\quad +\frac{\pi ^2 z^2 e^x (6 e^x+z^2)}{(4 e^x-z^2)^{7/2}}L'(x)-\frac{\pi ^2e^x (4 e^x+z^2)}{2 (4 e^x-z^2)^{5/2}} L''(x), \nn
\eta_3(x)&=\frac{2 \pi ^2 z e^{2 x} (2 e^{2 x}+9 e^x z^2+2 z^4) }{(4 e^x-z^2)^{9/2}}L(x)
\frac{ z e^x(4 e^{2 x}+10 e^x z^2+z^4)}{3 (4 e^x-z^2)^{7/2}} L^3(x) \\
&\quad-\frac{4\pi ^2 z e^{2 x}  (e^x+z^2) }{(4 e^x-z^2)^{7/2}}L'(x)
+\frac{2 \pi ^2 z e^{2 x}}{(4 e^x-z^2)^{5/2}} L''(x) ,\nn
t_3(x)&=\frac{ \pi  ze^x (32 e^{2 x}+98 e^x z^2+11 z^4) }{12 (4 e^x-z^2)^{7/2}}L(x)
+\frac{z e^x (2 e^x+z^2) }{3 \pi (4 e^x-z^2)^{5/2}}L^3(x) \nn
&\quad-\frac{ \pi  z e^x (16 e^x+11 z^2 ) }{6 (4 e^x-z^2)^{5/2}}L'(x)
-\frac{\pi  z (-28 e^x+z^2)}{24(4 e^x-z^2)^{3/2}} L''(x). \nonumber
\end{align}
Using these solutions, one can compute $I_{\pm,n}$ in \eqref{eq:Ipm}.
The coefficients of the small $N_f$ expansions \eqref{eq:Ipm-WKB} are given by
\be
\ba
I_{+,1}^{(1)}&=-\frac{\pi }{12 \left(4-z^2\right)^{3/2}}, \qquad 
I_{+,2}^{(0)}=\frac{\pi z^2}{4(4-z^2)^{5/2}},\qquad 
I_{+,3}^{(-1)}=-\frac{\pi  \left(4+z^2\right)}{4\left(4-z^2\right)^{5/2}}, \\
I_{-,1}^{(1)}&=\frac{z}{24 (4-z^2)}+\frac{1}{6(4-z^2)^{3/2}}\arctan \(\frac{z}{\sqrt{4-z^2}} \), \\
I_{-,2}^{(0)}&=-\frac{z}{24(4-z^2)^2}\left[ 8+z^2+\frac{12z}{\sqrt{4-z^2}} \arctan \(\frac{z}{\sqrt{4-z^2}} \) \right] , \\
I_{-,3}^{(-1)}&=\frac{z (28-z^2)}{24 (4-z^2)^2}+\frac{4+z^2}{2 (4-z^2)^{5/2}}\arctan \(\frac{z}{\sqrt{4-z^2}} \),
\ea
\ee
and
\be
\ba
I_{+,1}^{(2)}&=\frac{3 \zeta(3) \left(2+z^2\right)}{8 \pi  \left(4-z^2\right)^{5/2}},\qquad
I_{+,2}^{(1)}=\frac{\zeta(3)(2+z^2)}{8\pi(4-z^2)^{5/2}}, \qquad
I_{+,3}^{(0)}=\frac{\pi  \left(2+z^2\right)\log 2}{4 \left(4-z^2\right)^{5/2}}, \\
I_{-,1}^{(2)}&=-\frac{3\zeta(3)}{8\pi^2} \biggl[ \frac{3z}{(4-z^2)^2}+\frac{2(2+z^2)}{(4-z^2)^{5/2}}\arctan \(\frac{z}{\sqrt{4-z^2}} \) \biggr],\\
I_{-,2}^{(1)}&=-\frac{\zeta(3)}{8\pi^2} \left[ \frac{3z}{(4-z^2)^2}+\frac{2(2+z^2)}{(4-z^2)^{5/2}}\arctan \(\frac{z}{\sqrt{4-z^2}} \) \right],\\
I_{-,3}^{(0)}&=-\frac{\log 2}{4} \biggl[ \frac{3z}{(4-z^2)^2}+\frac{2(2+z^2) }{(4-z^2)^{5/2}}\arctan \(\frac{z}{\sqrt{4-z^2}} \) \biggr].
\ea
\ee

\section{'t Hooft expansion of the grand potential}\label{sec:'tHooft}
Here we compute the 't Hooft expansion of the grand potential.
As explained in \cite{MP2, GM}, the 't Hooft limit in the grand canonical ensemble corresponds to
\be
\mu \to \infty, \qquad N_f \to \infty, \qquad \h{\mu} =\frac{\mu}{N_f} \text{ : fixed}\,.
\ee
In this limit, the grand potential has the following ``genus'' expansion
\be
J(N_f,\mu)=\sum_{g=0}^\infty N_f^{2-2g} \mathcal{J}_g(\h{\mu}).
\ee
As noted in \cite{GM}, the genus zero contribution $\mathcal{J}_0(\h{\mu})$ is
given by the Legendre transformation of the planar free energy $F_0(\lambda)$.
Thus we have the relations
\be
\mathcal{J}_0(\h{\mu})=F_0(\lambda)-\lambda F_0'(\lambda),\qquad
\h{\mu}=-F_0'(\lambda).
\ee
The planar free energy was computed in \cite{GM}.
The result is expressed by the elliptic integral
\be
F_0''(\lambda)=-2\pi \frac{K(k)}{K(\sqrt{1-k^2})},\qquad
\h{\lambda}=\lambda+\frac{1}{8}=\frac{(1+k)^2}{8\pi^2}K^2(\sqrt{1-k^2}) ,
\ee
where $k$ is the elliptic modulus.
Using these relations, we find the large $\h{\mu}$ expansion of $\mathcal{J}_0(\h{\mu})$,
\begin{align}
\mathcal{J}_0(\h{\mu})&=
\frac{2}{3\pi^2}\h{\mu}^3-\frac{\h{\mu}}{8}+\frac{1}{\pi^2} \biggl[
\left(-\h{\mu}-\qu\right)e^{-4\h{\mu}}+\left(-4\h{\mu}^2+\qu\h{\mu}-\frac{7}{32}\right)e^{-8\h{\mu}} \nn
&\quad+\left(-\frac{128}{3}\h{\mu}^3+16\h{\mu}^2-\frac{46}{9}\h{\mu}+\frac{11}{27}\right)e^{-12\h{\mu}}\\
&\quad+\left(-\frac{2048}{3}\h{\mu}^4+\frac{1408}{3}\h{\mu}^3 
-178\h{\mu}^2+\frac{513}{16}\h{\mu}-\frac{2005}{768}\right)e^{-16\h{\mu}}\nn
&\quad+\left(-\frac{40960}{3}\h{\mu}^5+\frac{40960}{3}\h{\mu}^4-\frac{20096}{3}\h{\mu}^3
+1808\h{\mu}^2-\frac{20303}{75}\h{\mu}+\frac{6593}{375}\right)e^{-20\h{\mu}}+\cdots \biggr] .\nonumber
\end{align}
This should be compared with the worldsheet instanton correction \eqref{eq:WS-'tHooft}.

\end{document}